\documentclass[letterpaper,12pt]{article}
\usepackage{amssymb,latexsym,amsmath}

\usepackage{fullpage}
\usepackage{bbold}

\usepackage[pdftex]{graphicx}
\usepackage[table]{xcolor}
\usepackage{cite}

\makeatletter \@addtoreset{equation}{section}
\makeatother

\begin{document}

\begin{titlepage}
	\thispagestyle{empty}
	\begin{flushright}
		\hfill{DFPD-13/TH/07}\\
		\hfill{ROM2F/2013/04}
	\end{flushright}
	
	\vspace{35pt}
	
	\begin{center}
	    { \LARGE{\bf On the moduli space of spontaneously \\[4mm]
		broken $N=8$ supergravity}}
		
		\vspace{50pt}
		
		{F.~Catino$^{1}$, G.~Dall'Agata$^{2,3}$, G.~Inverso$^{4,5}$ and F.~Zwirner$^{2,3}$}
		
		\vspace{25pt}
		
		{
		$^1${\it  Institut de Th\'eorie des Ph\'enom\`enes Physiques, EPFL\\
		CH-1015 Lausanne, Switzerland}
		
		\vspace{15pt}
		
		$^2${\it  Dipartimento di Fisica e Astronomia ``Galileo Galilei''\\
		Universit\`a di Padova, Via Marzolo 8, 35131 Padova, Italy}
		
		\vspace{15pt}
		
	    $^3${\it   INFN, Sezione di Padova \\
		Via Marzolo 8, 35131 Padova, Italy}
		
		\vspace{15pt}
		
		$^4${\it Dipartimento di Fisica, Universit\`a di Roma ``Tor Vergata''\\
		Via della Ricerca Scientifica, 00133 Roma, Italy}
		
		\vspace{15pt}
		
		$^5${\it  INFN, Sezione di Roma ``Tor Vergata''\\
		Via della Ricerca Scientifica, 00133 Roma, Italy}
		}		
		
		\vspace{40pt}
		
		{ABSTRACT}
	\end{center}
	
	\vspace{10pt}
	
We analyze the moduli space of spontaneously broken $N=8$ supergravity theories in 4 dimensions with classical Minkowski vacua. We find that all the known classical vacua, as well as the several new ones we construct here, can be connected by sending some of the moduli to their boundary values. We also show that Cremmer--Scherk--Schwarz models can be viewed as special limits of more general CSO$^*$ gaugings, which allow for non-Abelian residual symmetries on the vacuum. Finally, we find that all the classical Minkowski vacua with fully broken supersymmetry found so far are unstable with respect to 1-loop corrections, which drive the effective potential to negative values.

\end{titlepage}

\baselineskip 6 mm



\section{Introduction} 
\label{sec:introduction}

The use of the embedding tensor formalism \cite{Nicolai:2000sc} has given new momentum to the analysis of the gauged versions \cite{deWit:2007mt} of maximal supergravity in 4 dimensions \cite{deWit:1977fk,Cremmer:1978ds,Cremmer:1979up}, which have recently revealed many new properties, both at the classical and at the quantum level.

A remarkable breakthrough has been the discovery of 1-parameter families of inequivalent models for certain choices of the gauge group \cite{Dall'Agata:2012bb}. 
This is especially striking in the case of the SO(8) model of \cite{deWit:1982ig}, which had been thoroughly examined in the past \cite{Warner:1983vz,Gunaydin:1983mi,deWit:1984va,Fischbacher:2009cj,Fischbacher:2010ki,Fischbacher:2010ec} and whose higher-dimensional origin is well understood in terms of a consistent truncation of 11-dimensional supergravity on the seven-sphere \cite{deWit:1983gs,deWit:1984nz,deWit:1986iy,Nicolai:2011cy,deWit:2013ija,Godazgar:2013nma}. 
One of the most intriguing aspects of the new SO(8)$_c$ theories is that they exhibit a vacuum structure \cite{Dall'Agata:2012bb}, \cite{ Borghese:2012qm,Kodama:2012hu,Borghese:2012zs,Borghese:2013dja} different from the one of the original SO(8) model. 
Hence they allow for new ways of breaking supersymmetry.
While all the details of the 4-dimensional action related to the appearance of the new parameter have now been worked out \cite{Dall'Agata:2012bb}, \cite{deWit:2013ija}, we still lack a string theory uplift, such as the one of the original model.

Another important recent development has been the introduction of an efficient technique to find vacua and compute their mass spectrum \cite{Inverso-thesis,Dibitetto:2011gm,DallAgata:2011aa}. 
This method relies on the old idea of simplifying computations on coset manifolds by evaluating physical quantities at the origin of the moduli space (see for instance \cite{Li:1986tk}), but combines it with the use of the embedding tensor. 
The result is a powerful technique, used both in $N=4$ \cite{Dibitetto:2011gm} as well as in $N=8$ supergravity \cite{Inverso-thesis}, \cite{DallAgata:2011aa}.
Not only many new vacua could be easily produced \cite{DallAgata:2011aa}, \cite{Dall'Agata:2012bb}, \cite{Borghese:2012qm,Kodama:2012hu,Borghese:2012zs,Borghese:2013dja}, but also simpler mass formulae have been derived, so that the problem of computing the classical spectrum can often be translated into a group-theoretical one \cite{DallAgata:2011aa},\cite{Kodama:2012hu}, \cite{Dall'Agata:2012cp}.
Furthermore, this same technique allowed to produce the first instance of a de Sitter vacuum of maximal supergravity for which slow-roll conditions are satisfied \cite{Dall'Agata:2012sx}.

Among all the models that can now be constructed and analyzed using this new technique, a particularly interesting class is given by those admitting classical Minkowski vacua with fully broken supersymmetry.
Before the introduction of this new formalism in \cite{Inverso-thesis,Dibitetto:2011gm,DallAgata:2011aa}, the only class of models with this property was the one proposed by Cremmer, Scherk and Schwarz (CSS) \cite{Cremmer:1979uq}.
These models have a positive semi-definite classical potential, whose minima are Minkowski vacua that break supersymmetry to $N=6,4,2$ or $0$ and where the overall scale of the gravitino masses is controlled by the classically undetermined expectation values of some moduli fields.
From the gauged supergravity point of view, these models can be realized by gauging a U(1) $\ltimes T^{24}$ group\footnote{We denote by $T^p$ the group of $p$ commuting translations and by $N^p$ any nilpotent group of dimension~$p$.} \cite{Andrianopoli:2002mf}, which is spontaneously broken to U(1).

Since \cite{DallAgata:2011aa}, we have a number of different models with classically stable Minkowski vacua, some of which also allow for non-Abelian residual gauge groups \cite{DallAgata:2011aa}, \cite{Kodama:2012hu,Borghese:2012zs}, \cite{Dall'Agata:2012sx}, \cite{Borghese:2013dja}.

In this paper we analyze in detail the structure and the properties of these new models, providing new examples.
We especially focus on the classical moduli space, finding that all of these models are interconnected.
We also address the issue of the 1-loop corrections to the scalar potential, extending the results of \cite{Sezgin:1981ac,Sezgin:1982pk,Gibbons:1984dg}, \cite{Dall'Agata:2012cp} and showing that these corrections destabilize all the known $N=0$ Minkowski vacua, from the old CSS models to the newly discovered ones.

To perform this study we heavily use the fact that the scalar manifold is the coset space E$_{7(7)}/$SU(8).
Because of this, we can relate the motion in the moduli space to the action of specific generators of the duality group on the embedding tensor and consider what happens at the boundary of the moduli space by taking appropriate limits.
We actually discuss, more generally, the procedure of contracting the gauge algebra in a way that consistently produces new models with Minkowski vacua, and relate it to the procedure described above.
By doing so, we show that the CSS gaugings with four parameters arise as special limits of a more general class of gaugings, which we call CSO$^*$ models, using the name of the gauge group.
These models have a non-Abelian residual gauge symmetry and again break supersymmetry to $N=6,4,2$ or $0$.
We should stress that, although prior to the introduction of this new technique many non-compact gaugings had been studied \cite{Hull:1984vg,Hull:1984ea,Hull:1984qz,Hull:1984rt,Hull:1984wa,Hull:1988jw,Hull:1984yy,Hull:2002cv}, including some of the CSO$^*$ models, no examples of Minkowski vacua with fully broken supersymmetry different from those of the CSS models had been discovered.

Another aspect we analyze in detail in this paper is the classical mass spectrum of spontaneously broken models on Minkowski vacua.
As already noted in \cite{Dall'Agata:2012cp}, most of the examples with residual Abelian factors in the gauge group have mass formulas that can be completely fixed in terms of the charge assignments of the various fields.
Here we will provide an interesting example that evades this simple rule, but for which we can also argue a mass formula.
These formulas are also interesting for another kind of analysis, namely for understanding the nature of the 1-loop corrections to the scalar potential.
In fact, recently two of us proved \cite{Dall'Agata:2012cp}, vastly extending previous results \cite{Sezgin:1981ac,Sezgin:1982pk,Gibbons:1984dg}, that \emph{any} spontaneously broken $N=8$ supergravity on a Minkowski vacuum has finite 1-loop corrections to the effective potential, thanks to new general supertrace identities obtained using the embedding tensor formalism, 
It was also found in \cite{Dall'Agata:2012cp} that in all these new examples the first non-trivial supertrace is Str$\,{\cal M}^8 > 0$ and that the 1-loop correction to the potential is always negative.

An interesting question we do not address in this paper is the higher-dimensional origin of the gauged supergravity models we analyze here.
Our analysis, however, allows us to identify a possible derivation of the 4-parameter CSS models.
For this reason, we added two appendices.
In Appendix~A we discuss how general CSS models may be obtained from M-theory, while in Appendix~B we present an $N=1$ truncation of the models of this paper that is especially convenient to produce fast checks of our results.


\section{Gaugings and contractions} 
\label{sec:gaugings_and_contractions}

In the following we will make use of two ingredients that are necessary to specify any gauged supergravity \cite{Cordaro:1998tx}, \cite{Nicolai:2000sc}: the symplectic frame and the embedding tensor. 
For this reason we now give a short review of their main features, also introducing some new results, to be used in the next sections.
In most of the text, we will be using `natural' units of gauged supergravity, where both the gauge coupling constant $g$ and the (reduced) Planck mass $M_P$ are set equal to one: in a few occasions, however, we will need to make the gauge coupling constant $g$ appear explicitly. 

Before introducing non-trivial gauge couplings, $N=8$ supergravity can be formulated in terms of an infinite number of different equivalent Lagrangians, which are not related to each other by local field redefinitions. 
Each of these Lagrangians depends on the choice of the symplectic frame, i.e.~on the choice of the embedding of the E$_{7(7)}$ duality group inside the Sp(56,$\mathbb R$) group that mixes electric and magnetic vector fields.
This choice also specifies the global symmetry group of the Lagrangian, which is a subgroup of E$_{7(7)}$.

Once the symplectic frame has been specified, so that 28 out of the ${56}$ vector fields $A_\mu^M$ have been declared fundamental, the embedding tensor $\Theta$ fully specifies the gauging and the corresponding Lagrangian and supersymmetry transformation rules.
In detail, $\Theta$ fixes the linear combinations of the 133 generators $t_{\alpha}$ of the E$_{7(7)}$ duality group that become generators of the gauge group:
\begin{equation}
	X_M = \Theta_M{}^\alpha \, t_{\alpha}.
\end{equation}
At the same time, $\Theta$ specifies what linear combinations of the vector fields appear in the gauge connection: $D_\mu = \partial_\mu - A^M_\mu \, \Theta_M{}^\alpha \, t_\alpha$.
As shown in \cite{deWit:2007mt}, locality and supersymmetry constrain the possible choices of $\Theta$ via the quadratic constraint
\begin{equation}
	\label{quad1}
  \Theta_{M}{}^{\alpha}\Theta_{N}{}^{\beta} \Omega^{MN} = 0
\end{equation}
and the linear constraints
\begin{equation}
	\label{lin}
  t_{\alpha M}{}^{N} \Theta_{N}{}^{\alpha} = 0 \, , \qquad (t_{\beta}t^{\alpha})_{M}{}^{N}\Theta_{N}{}^{\beta} = - \frac12 \, \Theta_{M}{}^{\alpha},
\end{equation}
where the index $\alpha$ in $t^\alpha$ has been raised with the inverse of the  ${\mathfrak e}_{7(7)}$ metric $\eta_{\alpha \beta}$.

Following \cite{Inverso-thesis}, \cite{DallAgata:2011aa}, we compute the scalar potential, the critical point conditions and the masses of the various fields at the point of the scalar manifold where all the scalars are vanishing, $\phi = 0$, in a frame where the scalar matrix ${\cal M}$ constructed in terms of the coset representatives $L$ trivializes, i.e.~${\cal M}(0) = L(0) L^T(0) = {\mathbb 1}$.
This means that we solve for $\Theta$ the linear and quadratic conditions (\ref{quad1})--(\ref{lin}), together with the critical point condition $\left.\partial_\rho \,V\right|_{\phi = 0} = 0$, where $\rho=1,\ldots, 70$ runs over the scalars associated to the non-compact generators of the E$_{7(7)}/$SU(8) coset, namely
\begin{equation}
	\Theta_{M}{}^\alpha [t_\rho]_{M}{}^{N}\Theta_N{}^\beta(\delta_\alpha^\beta + 7 \eta_{\alpha\beta})
			+\Theta_{M}{}^\alpha \Theta_M{}^\beta f_{\rho\alpha}{}^{\beta} = 0 \, ,
\end{equation}
where $f_{\alpha \beta}{}^\gamma$ are the $\mathfrak{e}_{7(7)}$ structure constants.

We then plug the result into the scalar potential
\begin{equation}
	\label{V0}
	V =\frac{1}{672} \, \Theta_{M}{}^\alpha \Theta_M{}^\beta\, (\delta_\alpha^\beta + 7\, \eta_{\alpha\beta}),
\end{equation}
 to obtain the cosmological constant, in 
\begin{equation}
	\label{mass matrix as isometry variation}
\begin{split}
	M_\rho{}^{\chi} \,=&\,\frac{1}{14}\left[\ \Theta_{M}{}^\alpha\, (t_\rho t^\chi)_M{}^N
\Theta_N{}^\beta\, (\delta_\alpha^\beta + 7\, \eta_{\alpha\beta}) +\Theta_{M}{}^\alpha\Theta_M{}^\beta f_{\rho \beta}{}^\gamma \eta^{\chi \sigma} f_{\sigma\gamma}{}^{\alpha} \right.\\[2mm]
&\left.+\ \Theta_{M}{}^\alpha [t_\rho]_{M}{}^{N} \Theta_N{}^\beta \eta^{\chi \sigma} f_{\sigma\beta}{}^{\alpha}
+\Theta_{M}{}^\alpha [t^\chi]_{M}{}^{N} \Theta_N{}^\beta f_{\rho\beta}{}^{\alpha} \right],
\end{split}
\end{equation}
with $t_{\rho}^T = t_{\rho}$ and $t_{\chi}^T = t_{\chi}$, to obtain the spectrum of the 70 scalar fields, and finally in 
\begin{equation}
	\label{vectormasses}
	M_{MN} = \frac{1}{24} \left[\Theta_M{}^\alpha \Theta_N{}^\beta\, (\delta_\alpha^\beta + \eta_{\alpha\beta})\right],
\end{equation}
to obtain the spectrum of the 56 vector fields\footnote{This new formula follows from the identification of the mass matrix of the vector fields with the square of the Killing vectors of the gauged isometries.} (where in the present formalism we will always have at least 28 vanishing eigenvalues, because we have only 28 physical vector fields).
Fermion masses are computed by first constructing the $A_1$ and $A_2$ tensors and then building the mass matrices out of them according to \cite{deWit:2007mt}.

In this work, we are primarily interested in gauged models that admit classical Minkowski vacua and in their moduli space.
For this reason we now recall an argument that can be used to identify such models \cite{DallAgata:2011aa}.
If $\Theta$ has a definite degree of homogeneity with respect to some non-compact generator $t \in \mathfrak e_{7(7)}$, which means that 
\begin{equation}
	(\delta_{t}\Theta)_M{}^\alpha = 
	 t_M{}^N \Theta_N{}^\alpha + t^\alpha{}_\beta\, \Theta_M{}^\beta
	 = k\, \Theta_M{}^\alpha,
\end{equation}
where $(t_{\gamma})^\alpha{}_{\beta} = f_{\gamma \beta}{}^\alpha$, then, for $k\neq 0$, the vacua of the corresponding model have vanishing cosmological constant.
This happens because the variation of the scalar potential with respect to such generator is proportional to the potential itself and therefore, for the first derivative to vanish, also the scalar potential has to vanish:
\begin{equation}\label{grading}
\delta_{t} V \propto 
  (\delta_{t}\Theta)_M{}^\alpha \Theta_M{}^\beta (\delta_\alpha^\beta + 7 \eta_{\alpha\beta})
  = k\, V = 0.
\end{equation}
Note that this also implies that the scalar field associated to $t$ is a modulus. 

While all the expressions given so far have been evaluated at $\phi = 0$, we can always recover the explicit field dependence by using the fact that each of the scalar fields can be associated to one of the generators of the E$_{7(7)}/$SU(8) coset.
For any given scalar field, we can define a scalar-dependent embedding tensor (from which we recover for instance the scalar potential) by parameterizing a geodesic in E$_{7(7)}/{\rm SU}(8)$ as $G(x) = e^{\, t \log x}$, with $x\in \mathbb R^*_+$, for some non-compact symmetric generator $t=t^T$ of the coset space. 
The corresponding field-dependent embedding tensor is then obtained by the appropriate action of the fundamental and adjoint representations of $G(x)$ on $\Theta$:
\begin{equation}
	\label{Thetax}
	[\Theta(x)]_M{}^\alpha\equiv \left[G(x)\Theta\right]_M{}^\alpha \equiv [G(x)]_M{}^N \, \Theta_{M}{}^\beta \, [G(x)]_\beta{}^\alpha.
\end{equation}

Having a field-dependent embedding tensor $\Theta(x)$, as in (\ref{Thetax}), is also extremely useful to construct new models starting from known ones, by taking appropriate limits and contractions.
Some CSO$(p,q,r)$ and CSO$^*(2p,2q)$ gaugings have been produced using a similar idea in \cite{Hull:1984qz}, \cite{Hull:2002cv}, but we will now show how to generalize those results by employing the embedding tensor formalism, following a method introduced in \cite{Fischbacher:2003yw} in the context of three-dimensional maximal supergravity.

The idea is to consider $\Theta(x)$ as a one-parameter deformation of the gauging described by $\Theta(x=1)$ and then take the limit $x \to 0$ to produce an inequivalent gauging.
In fact, for $x\in \mathbb R^*_+$, $\Theta(x)$ is still gauging a group isomorphic to the one defined by $\Theta(x=1)$. 
However, when $x\rightarrow0$, $G(x)$ becomes singular and usually the matrix $\Theta(x)$ diverges, because some of its entries are proportional to negative powers of $x$ in the limit $x \to 0$. 
Assuming that the most singular entries of $\Theta(x)$ are proportional to $x^{-p}$, we can cure the divergent terms by performing the limiting procedure together with a rescaling of the gauge coupling constant\footnote{When $\Theta$ has a non-vanishing degree of homogeneity $k$ with respect to some other generator $t'$, the redefinition of the gauge coupling constant is effectively obtained by acting on $\Theta$ with $\exp(t' \log y)$ and setting $y \sim x^{p/k}$ in the limit $x\to 0$. 
This is especially relevant when the original $\Theta$ satisfies the stationary point condition, because in that case $t'$ is a modulus.}
\begin{equation}
	g \rightarrow g'  \ x^p \, , 
\end{equation} 
using $g'$ as the new redefined gauge coupling constant:
\begin{equation}\label{contraction general definition}
g' \,  \Theta_{\rm contr} \equiv \lim_{x\rightarrow0} \left[g' \, x^p\ \Theta(x)\right]\,.
\end{equation}
The result is a new embedding tensor $\Theta_{\rm contr}$, which defines a gauge group that is generally not isomorphic to the original one.

We should note a few important consequences of this procedure.
First, since the action of $G(x)$ on $\Theta_{\rm contr}$ commutes with the limit in \eqref{contraction general definition}, we see that the contracted embedding tensor has a degree of homogeneity $-p$ with respect to the generator along which we performed the contraction:
\begin{equation}
G(x) \, \Theta_{\rm contr} = x^{-p} \,  \Theta_{\rm contr}.
\end{equation}
This means that for $p\neq0$ we obtain models that admit only Minkowski vacua or exhibit a runaway potential.
Then we also see that the action of $G(x)$ from the left may mix electric and magnetic vectors. 
This implies that, even if we start from an electric gauging in a given symplectic frame, the gauging described by the contracted embedding tensor may not be electric anymore in the same frame.
Obviously there may be instances where $G(x)$ does not introduce magnetic vectors in the linear combinations defining the new electric vectors, or it is such that they disappear in the contraction procedure, but the general case will bring outside the electric gaugings in the original frame.
Finally, while so far we treated $x$ as a parameter, as we shall discuss below in some \emph{interesting} cases $x$ can be identified with one of the moduli of a Minkowski vacuum, so that the limits $x \rightarrow0$ or $x \rightarrow +\infty$ correspond to approaching the boundary of the moduli space.
Moreover, by considering contractions along the moduli space, the vacuum condition is preserved.
Finally, if $x$ is the modulus corresponding to the generator along which we perform the contraction, it remains a modulus also in the resulting model.


\section{CSS gaugings} 
\label{sec:css_gaugings}

Until recently, the only example of a fully broken $N=8$ supergravity theory in 4 dimensions on a classically flat background was the CSS gauging of \cite{Cremmer:1979uq}.
The CSS model can be constructed in its electric frame by reducing maximal 5-dimensional supergravity on a circle and by twisting the reduction using the U-duality group \cite{Andrianopoli:2002mf}.
This procedure naturally selects a maximal subgroup of E$_{7(7)}$ that preserves E$_{6(6)}$, which corresponds to the duality group in 5 dimensions.
In fact, the electric frame for such models is the one following from the decomposition 
\begin{equation}
	\mathfrak e_{7(7)} = \mathfrak e_{6(6)} + \mathfrak{so}(1,1) + \mathbf{27}_{\mathbf{-2}} + \mathbf{27}'_{\mathbf{+2}},
\end{equation}
where $\mathbf{p_q}$ denotes the representations of $ [{\mathfrak e}_{6(6)}]_{\mathfrak{so}(1,1)} $, so that the fundamental representation of E$_{7(7)}$ decomposes as
\begin{equation}
	\label{decompo56}
	\mathbf{56} \rightarrow \mathbf{1}_{-\mathbf{3}} + {\mathbf{27}}^\prime_{-\mathbf{1}} + \mathbf{27}_{+\mathbf{1}}+ \mathbf{1}_{+\mathbf{3}},
\end{equation}
singling out the 28 electric and 28 magnetic vector fields.
The corresponding gauge algebra is
\begin{equation}
	\label{flatalgebra}
	[X_{28}, X_{\lambda}] = M_{\lambda}{}^\sigma \, X_{\sigma} \, ,
	\qquad \qquad
	[X_{\lambda}, X_{\sigma}] = 0 \, ,
\end{equation}
where $\lambda, \sigma = 1, \ldots, 27$.
In this explicit representation $X_{\lambda}$ is in the $\mathbf{27}'_{\mathbf{+2}}$ and $X_{28}$ is a generic Cartan generator of $\mathfrak{usp}(8) \subset \mathfrak e_{6(6)}$.
The gauge group is then a semidirect product of two Abelian factors
\begin{equation}
	\label{gaugegroupCSS}
	{\rm G = U(1)} \ltimes T^n, \qquad n \leq 24 \, , 
\end{equation}
and the matrix $M_{\lambda}{}^\sigma$ provides a representation of the U(1) $\subset $ USp(8) $\subset$ E$_{6(6)}$.
This fact constrains the matrix $M$, which depends on up to 4 real parameters $m_i$, $i=1,\ldots,4$, which determine the 4 Dirac masses of the gravitinos. Notice also the presence of at least 3 trivial U(1) vectors, with no fields charged with respect to them.

We now revisit this model using the language of the embedding tensor and discuss and clarify some of its features.

It is known that the Minkowski vacuum obtained by the CSS gauging preserves $8-2n$ supersymmetries, according to the number $n$ of non-zero mass parameters $m_i$.
The latter, in turn, are related to the eigenvalues of the matrix $M$, specifying the U(1) charges of the supergravity fields.
An interesting fact we will show in the following is that models with $2 \leq n \leq 4$ can be constructed by linear superposition of the embedding tensors defining models with a single supersymmetry breaking parameter.
This is a non-trivial statement, which does not apply to arbitrary gaugings, because in general linear combinations of two arbitrary embedding tensors do not fulfill anymore the consistency conditions.

In the electric frame discussed above, the generators of the group are
\begin{equation}
	\label{generatorsSS}
 X_\lambda = \begin{pmatrix}
 		{\mathbb0}_{27\times27}	 & -(M_\lambda)^\nu &  {\mathbb0}_{27\times27}		& \vec{0} 		      \\[3mm]
 		\vec{0}^T & 0 & \vec{0}^T & 0 \\[3mm]
 		M_\lambda{}^\sigma d_{\sigma\mu\nu} & \vec{0} & {\mathbb0}_{27\times27}								& \vec{0}\\[3mm]
 		\vec{0}^T		 & 0 & (M_\lambda)^\mu 		& 0 
 	   \end{pmatrix}
 ,\quad
 X_{28} = \begin{pmatrix}
	 		M_\mu{}^\nu	& \vec{0} & {\mathbb0}_{27\times27}	& \vec{0} \\[3mm]
 		\vec{0}^T & 0 & \vec{0}^T & 0 \\[3mm]
		{\mathbb0}_{27\times27}	 	& \vec{0} & -M_\nu{}^\mu	& \vec{0} \\[3mm]
 		\vec{0}^T & 0 & \vec{0}^T & 0 
 		  \end{pmatrix}\,,
\end{equation}
where $d_{\lambda\sigma\rho}$ is the $E_{6(6)}$ cubic invariant, whose normalization has been fixed so that $d_{\lambda\sigma\rho} = d^{\lambda\sigma\rho}$, $d_{\mu\lambda\sigma}
d^{\nu\lambda\sigma} = 10\, \delta_\mu^\nu $.
We now see explicitly that the scalar potential at the origin vanishes if and only if $M\in \mathfrak{usp}(8)$:
\begin{equation}
	V = \frac{1}{672}\left[{\rm Tr}(X_M X_M^T) + 7 \,{\rm Tr}(X_M X_M)\right] \propto {\rm Tr}[ M(M+M^T) ] = 0 .
\end{equation}
The origin is also a critical point of the potential because the variation of $V$ with respect to the 70 non-compact generators of $[{\rm E}_{6(6)}\times {\rm SO}(1,1)]\ltimes T^{27}$ vanishes.
We already know that the embedding tensor defining a flat group gauging has a non-trivial degree of homogeneity with respect to SO(1,1), because in the fundamental representation of E$_{7(7)}$ and in the electric frame we have 
\begin{equation}
 t_{SO(1,1)} = \text{diag}(-\mathbb1_{27},-3,\mathbb1_{27},+3) \, ,
\end{equation}
whose action on the potential gives just a multiplicative factor.
This implies that the vacua of this model always have a vanishing cosmological constant.
Variations of $V$ with respect to $T^{27}$ turn out to vanish because of the invariance of $d_{\lambda\sigma\gamma}$: 
\begin{equation}
	M_{(\lambda}{}^\rho d_{\sigma\gamma)\rho}=0.
\end{equation}
Finally, variations with respect to the non-compact generators in $E_{6(6)}$ give
\begin{equation}
 \delta_{E_{6(6)}}V \propto {\rm Tr}\left([t_{E_{6(6)}},M](M+M^T)\right)\,,
\end{equation}
which is also vanishing for $M\in \mathfrak{usp}(8)$.
The outcome of this analysis is that any CSS gauging is parameterized by a matrix $M\in \mathfrak{usp}(8)$ and has a Minkowski vacuum at the origin of the moduli space.

Since all  conditions discussed so far are linear in $M$, we can construct new consistent CSS gaugings by taking linear combinations of other CSS gaugings.
Moreover, since a linear combination of two matrices $M_1$ and $M_2$ defines a CSS gauging different from those defined by $M_1$ and $M_2$, generically it will break a different amount of supersymmetry.
Actually, also the mass matrix of the gravitini depends linearly on $M$, therefore
\begin{equation}
	A_1(M_1+M_2) = A_1(M_1) + A_1(M_2),
\end{equation}
which means that the number of supersymmetries preserved by the linear combination of $M_1$ and $M_2$ depends on the overlap of the supersymmetries of the two original models.
For instance, if we start with two $\mathcal N=2$ models described by
\begin{equation}
A_1(M_1)  = \text{blockdiag}(m_1\epsilon ,m_2\epsilon ,m_3\epsilon , {\mathbb0}_2),\quad 
A_1(M_2) = \text{blockdiag}({\mathbb0}_2,m'_2\epsilon ,m'_3\epsilon ,m'_4\epsilon),
\end{equation}
where $\epsilon= \left(\begin{smallmatrix}
0 & 1 \\ -1 & 0
\end{smallmatrix}\right)$, $A_1(M_1+M_2)$ has in general no vanishing eigenvalues and the model described by the linear combination $M_1+M_2$ has an $\mathcal N=0$ Minkowski vacuum. 

We stress that this simple superposition argument is not valid for any gauging.
In general, whenever we have two embedding tensors $\Theta_1$ and $\Theta_2$ that fulfill the consistency conditions and describe electric gaugings in the same symplectic frame, any linear combination of the two also fulfills identically the consistency conditions.
This happens because condition (\ref{lin}) is linear in $\Theta$, and (\ref{quad1}) is identically vanishing if the embedding tensors define electric gaugings in the same frame.
Although this tells us that any linear combination of $\Theta_1$ and $\Theta_2$ provides a consistent gauging, we cannot argue that the resulting model still admits a critical point at the origin, nor that such a point  has vanishing vacuum energy, because those conditions depend quadratically on $\Theta$.
So, in this respect the CSS gaugings are special.

\subsection{The moduli space of CSS gaugings} 
\label{sub:the_moduli_space_of_flat_gaugings}

We can now discuss the mass spectrum of the CSS models and their moduli space.
Most of the following discussion has already been given in \cite{Cremmer:1979uq}, \cite{Andrianopoli:2002mf}, \cite{Dall'Agata:2012cp}, but we will now put these results in the perspective of the present work, namely that of describing the moduli spaces of a wider class of spontaneously broken Minkowski vacua and discussing their stability against quantum corrections.

The spectrum of the generic 4-parameter CSS model is summarized in Table~\ref{tab_massesCSS}.
\begin{table}
	\renewcommand{\arraystretch}{1.5}\addtolength{\tabcolsep}{-1pt}%
\begin{center}
\rowcolors{1}{white}{gray!15}
\begin{tabular}{|c|c|}
\hline
	spin		& $m^2{}_{\text{ (multiplicity)}}$ \\
\hline\hline
	$2$ & $0_{(1)}$\\
 	$3/2$ & $e^{-2 \phi} m_i^2{}_{(2)}$ \\
	$1$ & $e^{-2 \phi} |m_i + m_j|^2_{(2)}$, \quad $e^{-2 \phi} |m_i - m_j|^2_{(2)}$,  \quad $0_{(4)}$\\
	$1/2$ & $0_{(8)}$, \quad  $e^{-2 \phi}m_i^2{}_{(4)}$,  \quad $e^{-2 \phi}|\pm m_i \pm m_j \pm m_k|^2_{(1)}$ \\
	$0$ & $0_{(30)}$, \quad $e^{-2 \phi} |m_i + m_j|^2_{(2)}$, \quad $e^{-2 \phi} |m_i - m_j|^2_{(2)}$, \quad  $e^{-2 \phi}|\pm m_1 \pm m_2 \pm m_3 \pm m_4|^2_{(1)}$\\\hline
\end{tabular}
\end{center}
\caption{Spectrum at the generic Minkowski critical point of a CSS model with parameters $m_i$. The lower numbers in round brackets denote the multiplicities not accounted for by the indices $i,j,k,l=1,2,3,4$, always taken in the order $i < j < k < l$. Goldstinos and Goldstone bosons providing  the additional degrees of freedom of massive gravitinos and vectors are formally included at zero mass.}
\label{tab_massesCSS}
\end{table}
It is clear that the residual supersymmetry varies according to the number of non-zero $m_i$ parameters.
Out of the 28 physical vector fields, 24 have non-zero masses and correspond to broken translations of the original gauge group, while 4 remain massless.
One of the massless vectors is the gauge boson of the residual non-trivial U(1) factor, while the remaining 3 massless vectors are simply inert and do not have any gauge interactions.
In fact, the gauge group is described by (\ref{gaugegroupCSS}) with $n=24$ when 3 or 4 mass parameters are non-zero, with $n=20$ when 2 mass parameters are non-zero and $n=12$ when only 1 mass parameter is non-vanishing (The number $n$ gets reduced if some of the mass parameter are equal to each other).
This also tells us that in the $N=0$ model and for generic values of the mass parameters, 24 of the massless scalars are actually Goldstone bosons and we are left with 6 real massless moduli fields.
All the masses in Table~1 have a non-trivial dependence on the moduli fields, but  it is simply an overall function $e^{-2 \phi}$.

The stability of the Minkowski vacuum with fully broken supersymmetry can be examined by considering the 1-loop effective potential, as a function of the moduli fields. 
It is known that the 1-loop effective potential can be expressed in terms of the supertraces of the field-dependent mass matrices
\begin{eqnarray}
{\rm Str} \, {\cal M}^{2k} & = & \sum_a (2J_a +1) \, (-1)^{2J_a} \,  [M^2_a(\phi)]^k \label{Supertrace}
\\ & = & 
{\rm Tr} \, [{\cal M}_{(0)}^2(\phi)]^k - 2 \ 	{\rm Tr} \, [{\cal M}_{(1/2)}^2(\phi)]^k + 3 \ {\rm Tr} \, [{\cal M}_{(1)}^2(\phi)]^k 
- 4 \ {\rm Tr} \, [{\cal M}_{(3/2)}^2(\phi)]^k \, , \nonumber
\end{eqnarray}
where $k=0,1,2,\ldots$, the index $a$ runs over the different particles in the spectrum, $M^2_a$  and $J_a$ are the corresponding squared-mass eigenvalues and spins.
It was already observed in \cite{Cremmer:1979uq}, \cite{Ferrara:1979fu} that in the model under consideration Str ${\cal M}^{2k} = 0$ for $k=0,1,2,3$ at $\phi = 0$.
The universal field-dependence of the mass spectrum makes it obvious that this remains true for any background value of the 3 complex moduli of the classical vacuum. 
The 1-loop contribution to the effective potential is then automatically finite (but field-dependent, a point overlooked in \cite{Cremmer:1979uq} but correctly identified in \cite{Andrianopoli:2002mf}) and its value is
\begin{equation}
\label{v1loop}
\Delta V_1 = \frac{1}{64 \, \pi^2} \, \sum_i (-1)^{2 J_i} \, (2J_i + 1) \, M_i^4 (\phi) \, \log M_i^2 (\phi) 
\equiv e^{-4\phi} \, f(m_1,m_2,m_3,m_4) \, . 
\end{equation}
It is easy to check that the function $f(m_1,m_2,m_3,m_4)$, explicitly defined by the above equation, vanishes for any of its four arguments going to zero, in agreement with the fact that all the supertraces identically vanish on a flat background if there is at least one unbroken supersymmetry. 
The other intriguing feature of $f(m_1,m_2,m_3,m_4)$, which emerges \cite{Dall'Agata:2012cp} from numerical inspection but we were unable to prove analytically, is the fact that it is negative semi-definite, and vanishes {\em only} in the supersymmetric limit discussed above.  

Before concluding this discussion of the CSS models, we would like to add some comments on their higher-dimensional origin.
As mentioned above, CSS models can be obtained in the electric frame by reducing 5-dimensional supergravity and introducing a non-trivial Scherk--Schwarz twist using the $U$-duality group of the 5-dimensional theory.
This allows to reproduce the most general CSS gauging with up to four non-zero mass parameters $m_i$.
It is known how to reproduce a similar result from $M$-theory reductions on twisted tori \cite{Cremmer:1979uq, Dall'Agata:2005ff,D'Auria:2005dd,Dall'Agata:2005fm,D'Auria:2005rv}, but it is also known that such geometric reductions can only produce at most 3 of the 4 mass parameters of the CSS models.
By a simple comparison of the potential terms of $N=8$ supergravity  with those generated by flux compactifications of $M$-theory, assuming a framework where the starting point defining the electric frame of the 4-dimensional theory is the compactification on a torus, we can identify the fourth mass parameter with a non-geometric flux: $\theta_{77}$.
Since this is not directly related to the main scope of this work, we refer the reader to Appendix~A, where we give some more details on the identification.



\section{CSO$^*$ gaugings} 
\label{sec:cso_gaugings}

In this section we introduce the CSO$^*$ models and discuss their vacua and moduli space.
Models with such gauge groups as electric subgroups of the SU$^*(8)$ subgroup of E$_{7(7)}$ have been also discussed in \cite{Hull:2002cv} .
Here, however, we will give a more general construction, following the argument that we may have many inequivalent models with the same gauge group, as in \cite{Dall'Agata:2012bb}, and present many new vacua in addition to the $N=2$ Minkowski vacuum of the CSO$^*(6,2)$ model discussed in \cite{Hull:2002cv}.
Actually, we will show how the CSO$^*$ models provide a non-Abelian generalization of the CSS models, which appear as limiting cases at the boundary of the moduli spaces of the Minkowski vacua of CSO$^*$ models.

\subsection{CSO* definition and relevant symplectic frames} 
\label{sub:cso_groups_and_their_embedding}

The CSO$^*(2p,8-2p)$ groups ($p=0,1,2,3$) are defined as the set of complex matrices $M$ satisfying
\begin{equation}
	M \, \Omega = \Omega \, M^* \, , 
	\qquad
	\qquad
	M^T \, \eta \,  M = \eta \, ,
\end{equation}
where 
\begin{equation}
	\Omega = \left(\begin{array}{cc}
	{\mathbb 0} & {\mathbb 1}_4 \\
	-{\mathbb 1}_4 & {\mathbb 0}
	\end{array}\right), \qquad \qquad \eta = \left(\begin{array}{cc}
	{\mathbb 1}_{2p} & 
	{\mathbb 0} \\
	{\mathbb 0} & {\mathbb 0}_{8-2p}
	\end{array}\right).
\end{equation}
More precisely, they are defined as contractions of SO$^*(8)$, so that the SU$^*(8-2p)$ factor that can be seen to satisfy the above conditions is not gauged.
At the algebra level:
\begin{equation}
	\mathfrak{so}^*(8) = \mathfrak{so}(6,2), \quad 	\mathfrak{so}^*(6) = \mathfrak{su}(3,1), \quad	\mathfrak{so}^*(4) = \mathfrak{so}(3)+\mathfrak{so}(2,1),\quad 	\mathfrak{so}^*(2) = \mathfrak{so}(2).
\end{equation}
As mentioned above, such gauge groups have been discussed in \cite{Hull:2002cv} as deformations of $N=8$ supergravity in the SU$^*$(8) frame, which is one of the possible frames determined by maximal subgroups of E$_{7(7)}$.
However, we will find models with the same gauge groups arising as deformations of the ungauged theory in the standard SL($8,{\mathbb R}$) frame, by introducing the 1-parameter family of inequivalent deformations of the SO(6,2) $\simeq$ SO$^*(8)$ model, which can also be naturally embedded in SL($8,{\mathbb R}$).
For this reason, we first review the SO$^*(8)$ model as constructed in \cite{DallAgata:2011aa}, \cite{Kodama:2012hu} in the framework of the new $c$-deformed supergravities of \cite{Dall'Agata:2012bb} and then use the $c=1$ model, which has a Minkowski vacuum, to generate the others by contractions.

Since in what follows we need to use different symplectic frames, we briefly discuss their relation \cite{deWit:2002vt}.
The standard formulation of ungauged $N=8$ supergravity \cite{deWit:1982ig}, which has also been used as a basis to build many CSO gaugings \cite{Cordaro:1998tx}, is given in the so-called SL$(8,{\mathbb R})$ frame. 
In this frame the Lagrangian is invariant under SL$(8,{\mathbb R})$ and the gauge fields transform in the $\mathbf{28}+\mathbf{28}'$ representation: $A_\mu^M = \{A_\mu^{[AB]}, A_{\mu\,[AB]}\}$, where $A,B=1,\ldots,8$ are indices labeling the fundamental representation of $\mathfrak{sl}$(8,${\mathbb R}$). 
Also the 133 generators of the E$_{7(7)}$ group in the SL$(8,{\mathbb R})$ basis can be decomposed as  $\mathbf{133} \to \mathbf{63} + \mathbf{70},$ where the first 63 are the generators of the SL($8,{\mathbb R}$) subgroup of E$_{7(7)}$, which we name $t_A{}^B$, and the remaining 70 are described by a rank 4 totally antisymmetric tensor $t^{ABCD}$.
CSS gaugings are electric in the E$_{6(6)}$ basis instead.
Following (\ref{decompo56}), the $\mathfrak{e}_{7(7)}$ elements that have an electric action in this basis have the form \cite{deWit:2002vt}
\begin{equation}
	\rm{E}_{6(6)} = \left(\begin{array}{cccc}
	K_{27} &  &  & \\
	 & 0 &  & \\
	 &  & -K_{27}^T & \\
	  &  &  & 0
	\end{array}\right), \quad t_{SO(1,1)} = \left(\begin{array}{cccc}
	-{\mathbb 1}_{27} &  &  & \\
	 & -3 &  & \\
	  &  & {\mathbb 1}_{27} & \\
	   &  &  & 3
	\end{array}\right), 
\end{equation}
where $K_{27} \in {\mathfrak{e}_{6(6)}}$ are in the representation $\mathbf{27}$, and 
\begin{equation}
	\mathbf{27}'_{+2} = \left(\begin{array}{cccc}
	 {\mathbb 0}_{27}& \vec{t} &  & \\
	  & 0 &  & \\
	  d_{\lambda \sigma \gamma} t^\gamma &  & {\mathbb 0}_{27} & \\[2mm]
	   &  & -\vec{t}^{\ T} & 0
	\end{array}\right),
\end{equation}
where $t^\gamma$ is a 27-dimensional vector of parameters.
We can understand how to go from one frame to the other by analyzing the common subgroup SL($2,{\mathbb R}$) $\times $ SL($6,{\mathbb R}$) $ \times $ SO(1,1).
The SL$(8,{\mathbb R})$ representations for the vector fields decompose as
\begin{eqnarray}
	\mathbf{28} &\to& (\mathbf{2},\mathbf{6})_{-1} + (\mathbf{1},\mathbf{15})_{+1} + (\mathbf{1},\mathbf{1})_{-3}, \\[2mm]
	\mathbf{28}' &\to& (\mathbf{2},{\mathbf{6}'})_{1} + (\mathbf{1},\mathbf{15}')_{-1} + (\mathbf{1},\mathbf{1})_{+3},
\end{eqnarray}
while 
\begin{equation}
	\mathbf{27}'_{-1} + \mathbf{1}_{-3} \to (\mathbf{2},\mathbf{6})_{-1} + (\mathbf{1},\mathbf{15}')_{-1} + (\mathbf{1},\mathbf{1})_{-3}.
\end{equation}
This means that to change frame we have to exchange 15 magnetic vectors of SL($8,{\mathbb R}$) with 15 electric ones.
Finally, the SU$^*(8)$ frame requires the electric vector fields to transform in the representation $\mathbf{28}$ of SU$^*(8)$.
This can be achieved by acting on the E$_{6(6)}$ frame with the matrix \cite{deWit:2002vt}
\begin{equation}
	\label{rotation to SUstar}
	E_{E_6 \to SU^*(8)} = \frac{1}{\sqrt2}\left(\begin{array}{cccc}
	{\mathbb 1}_{27} &  & -{\mathbb 1}_{27} & \\
	 & 1 &  & 1\\
	{\mathbb 1}_{27} &  & {\mathbb 1}_{27} & \\
	 & -1 &  & 1
	\end{array}\right),
\end{equation}
therefore mixing all electric and magnetic vectors of the E$_{6(6)}$ and/or SL($8,{\mathbb R}$) frames.
For our following discussion it is also important to note that the combined action of Sp(56$,{\mathbb R}$) bringing from the SL(8,${\mathbb R}$) to the SU$^*(8)$ frame is given by the following matrix:
\begin{equation}
	\label{SLtoSU*}
	E_{SL(8,{\mathbb R}) \to SU^*(8)} = \frac{1}{\sqrt2}\left(\begin{array}{cccccc}
	{\mathbb 1}_{12} && &   -{\mathbb 1}_{12} & & \\
	& {\mathbb 1}_{15}& && {\mathbb 1}_{15} & \\
	 & & 1 &  & &1\\
	{\mathbb 1}_{12}& & &  {\mathbb 1}_{12}& & \\
	& -{\mathbb 1}_{15}& && {\mathbb 1}_{15} & \\
	 & &-1 &  && 1
	\end{array}\right).
\end{equation}


\subsection{SO$^*(8)$ and its moduli space} 
\label{sub:so_8_}

In the SL($8,{\mathbb R}$) frame we can gauge a CSO($p,q,r$) group by choosing the embedding tensor as
\begin{equation}
	\label{thetacoupling}
	\Theta_{M}{}^\alpha = \Theta_{AB}{}^C{}_D \propto \delta_{[A}^C \theta^{\vphantom{C}}_{B]D},
\end{equation}
where $\theta_{AB}$, which couples the electric vectors to the SL(8,${\mathbb R}$) generators $t_C{}^D$, is chosen to be proportional to the CSO($p,q,r$) metric \cite{Cordaro:1998tx}, \cite{deWit:2007mt}:
\begin{equation}
	\theta = \left(\begin{array}{ccc}
	{\mathbb 1}_p &  & \\
	& -{\mathbb 1}_q & \\
	& & {\mathbb 0}_r
	\end{array}\right)\,.
\end{equation}
Following \cite{DallAgata:2011aa}, when $r=0$, we can gauge the same model also by introducing a second tensor $\xi$ so that
\begin{equation}
	\label{xicoupling}
	\Theta^{AB\,C}{}_D \propto \delta^{[A}_D \xi_{\vphantom{D}}^{B]C},
\end{equation}
and 
\begin{equation} 
	\label{xitheta}
	\xi = c\, \theta^{-1},
\end{equation} 
where $c$ is a real parameter and the inverse is needed because of the different transformation properties of $\xi$ with respect to SL$(8,{\mathbb R})$.
This produces inequivalent gaugings in a definite range, to be determined for each gauge group\footnote{These same deformations have been sometimes labelled by a parameter $\omega$. 
The conversion is given by $c=\tan\omega$ up to a rescaling of the gauge coupling constant \cite{Dall'Agata:2012bb}.} (for SO(8)$_c$ models $c \in [0,\sqrt2-1]$, for SO(6,2)$_c$ models we expect $c \in [0,1]$).
As noted in \cite{Dall'Agata:2012bb}, \cite{Borghese:2012qm,Kodama:2012hu,Borghese:2012zs,Borghese:2013dja}, \cite{Dall'Agata:2012sx}, \cite{deWit:2013ija}, varying $c$ also varies the structure of the scalar potential and the number of critical points.
For the SO(6,2)$_c$ models, the analysis of \cite{DallAgata:2011aa} shows that there is a Minkowski vacuum at $c=1$, which disappears for $c\neq 1$ (which explains why it was not found in the $c=0$ model discussed in \cite{Hull:1984qz},\cite{Hull:2002cv}).
Although most of the details of this model have been worked out in \cite{DallAgata:2011aa}, \cite{Kodama:2012hu}, we will now review and extend its discussion in order to use it as a basis for the following developments.

First of all, we must clarify the relation between the SL(8) and SU$^*$(8) symplectic frames for what concerns the SO$(6,2)\simeq{\rm SO}^*(8)$ gauging. 
The discussion at the end of the previous section and Eq.~\eqref{SLtoSU*} in particular show that the SO$(6,2)_{c=1}$ gauging becomes electrical when we switch to the SU$^*$(8) frame.
Hence we can say that the SO$^*$(8) model defined in \cite{Hull:2002cv} in the SU$^*(8)$ frame corresponds in our language to SO(6,2)$_{c=1}$, and it admits a Minkowski vacuum as proved in \cite{DallAgata:2011aa}. 
Of course also all the other SO$(6,2)_c$ gaugings can be rotated to the SU$^*(8)$ frame, but they still contain magnetic vectors in the gauge connection\footnote{Also note that while the CSO$(p,q,r)$ contractions of SO(6,2) can only be defined when $c=0$, the CSO$^*(2p,8-2p)$ gaugings can be obtained as contractions only from the SO$^*(8)$ model that is electric in the SU$^*(8)$ frame, which in our language translates to the condition $c=1$. 
This is a consequence of the quadratic constraint on the $\bf 36$ and $\bf 36'$ representations of either SL(8$,{\mathbb R}$) or SU$^*(8)$, in which the embedding tensor of the SO(6,2) models sits.
For different values of $c$ we may reach new contractions, along the lines of those defined for $\theta \xi=0$ \cite{DallAgata:2011aa}. We discuss them in Section \ref{sub:extra_sustar8_models}.}.

The [SO$^*(8)\simeq$ SO$(6,2)]_{c=1}$ model has a Minkowski vacuum with a residual ${\rm SO(6)\times U(1)}$ gauge group and no unbroken supersymmetries.
The spectrum at this critical point arranges in representations of ${\rm SO(6)\times U(1)}$ as shown in Table~\ref{tab_massesso62}.
Curiously, the spectrum is identical to the one of the CSO(2,0,6) gauging, which coincides with the one of the CSS$_{N=0}$ model where $m_1=m_2=m_3=m_4$.

\begin{table}
	\renewcommand{\arraystretch}{1.3}\addtolength{\tabcolsep}{-1pt}%
\begin{center}
\rowcolors{1}{white}{gray!15}
\begin{tabular}{|c|c|l|}
\hline
	spin		& $m^2{}_{\text{ (SO(6) representation, U(1) charge)}}$ \\
\hline\hline
	$2$ & $0_{(\mathbf{1},0)}$\\
 	$\frac32$ & $\frac{1}{8}_{(\mathbf{4},1)}$, \ $\frac{1}{8}_{(\overline{\mathbf{4}},-1)}$ \\
	$1$ & $0_{(\mathbf{1},0)}$, \ $0_{(\mathbf{15},0)}$, \ $\frac12_{(\mathbf{6},2)}$, \ $\frac12_{(\mathbf{6},-2)}$ \\
	$\frac12$ & $0_{(\mathbf{4},1)}$, \ $0_{(\overline{\mathbf{4}},-1)}$, \ $\frac18_{(\textbf{20},1)}$, \ $\frac18_{(\overline{\textbf{20}},-1)}$, \ $\frac98_{(\textbf{4},-3)}$, \ $\frac98_{(\overline{\textbf{4}},+3)}$\\
	$0$ & $0_{(\mathbf{1},0)}$, \ $0_{(\mathbf{15},0)}$, \ $0_{(\mathbf{6},2)}$, \ $0_{(\mathbf{6},-2)}$, \ $0_{(\mathbf{20}',0)}$, \ $\frac12_{(\mathbf{10},-2)}$, \ $\frac12_{(\overline{\mathbf{10}},2)}$, \ $2_{(\mathbf{1},4)}$, \ $2_{(\mathbf{1},-4)}$
	\\\hline
\end{tabular}
\end{center}
\caption{Spectrum of the [SO$^*(8)\simeq$ SO$(6,2)]_{c=1}$ model at the Minkowski critical point with $\rm G_{res} = SO(6)\times SO(2)$.}
\label{tab_massesso62}
\end{table}

Analyzing Table~\ref{tab_massesso62} we see that all the gravitinos are massive, as expected for a Minkowski vacuum with fully broken supersymmetry, while 16 vector fields are massless and sit in the adjoint representation of the residual gauge symmetry group.
We also have 8 formally massless fermions that play the role of the goldstinos and 48 formally massless scalar fields.
As it is clear from the representations, 12 of these massless scalars are indeed Goldstone bosons of the broken gauge symmetry, while the remaining 36 massless fields are real moduli or would-be Goldstone bosons associated with the possible further breaking of the residual gauge symmetry group.
In fact, by giving a non-trivial expectation value to some of these fields we can further break the residual gauge group to U(1)$^4$.
When this happens, some of the massless fields become massive and we are left with a total of 6 massless scalars, which parameterize a ${\rm \left[SU(1,1)/U(1)\right]^3}$ moduli space.
Obviously, the way this moduli space is embedded in the original E$_{7(7)}/$SU(8) depends on the way we break SO(6) to U(1)$^3$ and this explains why the number of massless fields at the maximally symmetric point is higher than the one along the various branches we now analyze.
However, in the SL($8,{\mathbb R}$) frame, we can use the residual gauge symmetry to fix the generators of the U(1)$^4$ to be
\begin{equation}
	t_1{}^2 - t_2{}^1, \qquad t_3{}^4 - t_4{}^3, \qquad t_5{}^6 - t_6{}^5, \qquad t_7{}^8 - t_8{}^7.
\end{equation}
The 6 moduli fields correspond to generators of E$_{7(7)}$ that are neutral with respect to these U(1)$^4$ and the remaining moduli space ${\rm \left[SU(1,1)/U(1)\right]^3}$ can be parameterized by vielbeins extracted from the Maurer--Cartan forms obtained from the coset representatives
\begin{align}
X_i \equiv \exp(\ell_i\,\log x_i),\qquad E_i \equiv \exp(\lambda_i\,\log e_i),\qquad i=1,2,3,
\end{align}
where $\ell_i = \ell^T_i,\ \lambda_i=\lambda^T_i$ are associated to
\begin{eqnarray}
	\ell_1 &=& t_1{}^1+t_2{}^2-t_7{}^7-t_8{}^8, \\[2mm]
	\ell_2 &=& t_3{}^3+t_4{}^4-t_7{}^7-t_8{}^8, \\[2mm]
	\ell_3 &=& t_5{}^5+t_6{}^6-t_7{}^7-t_8{}^8, \\[2mm]
	\lambda_1 &=& t^{1278} + t^{3456}, \\[2mm]
	\lambda_2 &=& t^{3478} + t^{1256}, \\[2mm]
	\lambda_3 &=& t^{5678} + t^{1234}. 
\end{eqnarray}
It is interesting to note that this moduli space is the same as the one of the STU model.
However, differently from what happens for the analogous truncation of the SO(8)$_c$ models, this truncation has a potential that depends on the $c$ parameter.
The full scalar potential and/or the dependence of the mass formulas on these moduli can be obtained by acting with the above generators on the embedding tensor of the SO$^*(8)\simeq{\rm SO}(6,2)_{c=1}$, namely $\Theta_0^{\mathfrak{so}^*(8)}$, which is defined by $\theta = \xi = {\rm diag}\{1,1,1,1,1,1,-1,-1\}$.
The three factors commute with each other, but since $[\ell_i, \lambda_i] \neq0$, we need to fix the order in which they act on $\Theta_0^{\mathfrak{so}^*(8)}$:
\begin{equation}
	\label{ThetaSO8*}
	\Theta^{\mathfrak{so}^*(8)}(x_i,\,e_i)\equiv\prod_{i=1}^3 (X_i\, E_i) \ \Theta_0^{\mathfrak{so}^*(8)}.
\end{equation}
Of course, any other ordering or parametrization of the coset space is equivalent up to a change of coordinates.

The general mass spectrum follows the rule outlined in \cite{Dall'Agata:2012cp}, which means that we can express it in terms of a general mass formula
\begin{eqnarray}
\rm{spin}\ 2: & & M^2 = 0 \, , 
\nonumber \\[2mm]
\rm{spin}\ \frac32 : & & M_a^2 = \left( \vec{q}_a \right)^2 \, , \quad a=1,\ldots,8,
\nonumber \\[2mm]
\rm{spin}\ 1 : & & M_{ab}^2 = \left( \vec{q}_a + \vec{q}_b \right)^2  \, , \quad 1\leq a < b \leq 8,\label{U1spectrum}\\[2mm]
\rm{spin}\ \frac12: & & M_{abc}^2 = \left( \vec{q}_a + \vec{q}_b + \vec{q}_c \right)^2  \, , \quad 1\leq a < b < c \leq 8,
\nonumber \\[2mm]
\rm{spin}\ 0 : & & M_{abcd}^2 = \left( \vec{q}_a + \vec{q}_b + \vec{q}_c + \vec{q}_d \right)^2 \, , \quad 1\leq a <b<c<d\leq 8,
\nonumber
\end{eqnarray}
where
\begin{equation}
(\vec{Q})^2 = \sum_{\alpha=1}^4 Q^\alpha \, Q^\alpha\ \mu_{\alpha}^2 \, .
\label{U1metric}
\end{equation}
The introduction of the mass dependence on the full moduli space, however, changes the values we have to take for $\vec q$ with respect to those given in \cite{Dall'Agata:2012cp}.
In detail, 
\begin{eqnarray}\label{mu1}
\mu_1^2 &=& \frac{(x_1^2-x_2^2)(x_1^2-x_3^2)(1+x_2^2x_3^2)}{8 x_1^2 x_2^2 x_3^2}, \\[2mm]
\mu_2^2 &=& \frac{(x_1^2-x_2^2)(x_3^2-x_2^2)(1+x_1^2x_3^2)}{8 x_1^2 x_2^2 x_3^2}, \\[2mm]
\mu_3^2 &=& \frac{(x_1^2-x_3^2)(x_2^2-x_3^2)(1+x_1^2x_2^2)}{8 x_1^2 x_2^2 x_3^2}, \\[2mm]
\mu_4^2 &=& \frac{(1+x_1^2x_2^2)(1+x_1^2x_3^2)(1+x_2^2x_3^2)}{8 x_1^2 x_2^2 x_3^2}, \label{mu4}
\end{eqnarray}
and 
\begin{eqnarray}
\vec{q}_1 = - \vec{q}_2 = \frac{e_2 e_3}{e_1}(+1,+1,+1,+1) \, , 
& &                      
\vec{q}_3 = - \vec{q}_4 = \frac{e_1 e_3}{e_2}(+1,+1,-1,-1) \, , 
\nonumber \\
\vec{q}_5 = - \vec{q}_6 = \frac{e_1 e_2}{e_3}(+1,-1,+1,-1) \, , 
& & 
\vec{q}_7 = - \vec{q}_8 = \frac{1}{e_1e_2e_3}(+1,-1,-1,+1) \, .
\label{so62charges}
\end{eqnarray}
We stress that now the `charge vectors' $\vec{q}$ are no longer constants as in Ref.~\cite{Dall'Agata:2012cp}, but also field-dependent, through the moduli $(e_1,e_2,e_3)$.   

An interesting point that we can now address is the existence of regions in the moduli space without tachyonic scalars.
Unless all of the $x_i$ moduli have the same value, at least one of the $\mu_1,\mu_2,\mu_3$ parameters is negative and some of the scalar squared masses can become negative.
We can also see, however, that there are regions where all the scalar squared masses are positive and the moduli have to change by a finite amount to generate tachyonic modes.
One such region appears at $x_1=x_3=e_2=e_3=1$, for $e_1 \neq 1$ and also $x_2 \neq 1$.
For all such values the spectrum does not contain tachyons.
Also, whenever one of the masses is much smaller than the others, the spectrum of the model approaches that of a supersymmetric one and therefore the unstable region shrinks accordingly.
Although we did not perform an exhaustive analysis, we could see that there are stable vacua also in regions of  finite volume where all the moduli get a non-trivial expectation value.

Another interesting point is that the 4 gravitino masses can be rescaled independently by tuning the values of the moduli:
\begin{eqnarray}
	M_1 &=& \frac{e_2 e_3}{e_1}\,\frac{\sqrt{(1+x_1^4)(1+x_2^4)(1+x_3^4)}}{2\sqrt2\,x_1 x_2 x_3}\,,\\[2mm]
	M_2 &=& \frac{e_1 e_3}{e_2}\,\frac{\sqrt{(1+x_1^4)(1+x_2^4)(1+x_3^4)}}{2\sqrt2\,x_1 x_2 x_3}\,,\\[2mm]
	M_3 &=& \frac{e_1 e_2}{e_3}\,\frac{\sqrt{(1+x_1^4)(1+x_2^4)(1+x_3^4)}}{2\sqrt2\,x_1 x_2 x_3}\,,\\[2mm]
	M_4 &=& \frac{1}{e_1e_2e_3}\,\frac{\sqrt{(1+x_1^4)(1+x_2^4)(1+x_3^4)}}{2\sqrt2\, x_1 x_2 x_3}\,.
\end{eqnarray}                 
This means that we can always send some of them to zero by moving towards the boundary of the moduli space, enhancing the number of supersymmetries of the vacuum, from $N=0$ to $N=2,4$ or $6$.
Effectively, we can treat the 4 masses as moduli and this is different from the CSS models, where the gravitino masses are determined by 4 parameters, which, however, rescale all in the same fashion by changing the values of the moduli fields.
Note moreover that, as long as we only vary the $e_i$ moduli, keeping $x_i=1$, the full mass spectra coincide with those of the CSS model with the same gravitino mass parameters.


\subsection{Going to the boundary} 
\label{sub:limits_in_moduli_space}

As promised, we now consider what happens when we move in the moduli space towards its boundary.
Obviously, the contraction procedure defined in \eqref{contraction general definition} can be applied for any generic direction in E$_{7(7)}/{\rm SU}(8)$.
However, it is interesting to see how the different Minkowski models we know are connected to each other when performing singular limits along the moduli space, hence preserving not only the embedding tensor constraints but also the vacuum condition.
While doing so, we will see that several new models arise, too, and we will unveil an unexpected and intriguing link between the CSO$^*$ and CSS gaugings.

The starting point is the embedding tensor (\ref{ThetaSO8*}), which is a function of 6 parameters, corresponding to the 6 massless moduli of the SO$^*(8)$ gauged model.
We then apply the procedure outlined in section 2, by taking singular limits to the boundary of the moduli space (rescaling the gauge coupling constant accordingly) and obtain new models.
To identify the resulting gauge algebra, it is sometimes sufficient to check the rank of $\Theta$ and the signature of the gauge invariant metric $\eta_{MN} \equiv \Theta_M{}^\alpha \Theta_N{}^\beta \eta_{\alpha\beta}$, otherwise we can always resort to computing the structure constants explicitly.

Since we have six moduli, three $x_i$ and three $e_i$, we can approach the boundary in several different directions, sending some of their combinations to zero.
We summarize the outcome of this procedure in Table~\ref{tab:contractions}, where we give the gauge group corresponding to the embedding tensor obtained by the contractions with respect to the $x_i$ and $e_i$ moduli.
Taking singular limits in any other combination always reproduces one of the gauge groups\footnote{Reaching the same gauge group through different combinations of contractions generally gives embedding tensors that are not identical to each other.
However, the fact that all the physical properties that we can compute match strongly suggests that these embedding tensors are equivalent up to a duality and do not represent inequivalent theories.} in Table~\ref{tab:contractions}.
Also the mass spectra coincide up to a reordering of the moduli. 
Moreover, we expect the $\mathrm U(1)^2\ltimes N^r$ families of gaugings to be related to the two-fold Scherk--Schwarz reductions discussed in \cite{deWit:2007mt}.

\begin{table}
	\renewcommand{\arraystretch}{1.3}\addtolength{\tabcolsep}{-1pt}%
\begin{center}
\rowcolors{1}{white}{gray!15}
\begin{tabular}{|r|llll|}
\hline
& & $x_1\rightarrow0$ & $x_1, x_2\rightarrow0$ & $x_1, x_2, x_3\rightarrow0$\\
\hline\hline
& SO$^*(8)$ & $({\rm SO}(4)\times {\rm SO}(2,2))\ltimes T^{16}$ & $[{\rm U}(1)^2\ltimes N^{26}]_{{N} = 0}$ & ${\rm CSS}_{ N=0}$ \\[.2em]
$e_1^{-1}, e_2, e_3\rightarrow0$ & ${\rm CSO}^*(6,2)$ & $({\rm SO}^*(4)\times {\rm U}(1))\ltimes {N^{20}}$ & $[{\rm U}(1)^2\ltimes N^{24}]_{{N} = 2}$ & ${\rm CSS}_{ N=2}$ \\[.2em]
$e_3\rightarrow0$ & ${\rm CSO}^*(4,4)$ & $[{\rm U}(1)^2 \ltimes N^{20}]_{{N} = 4}$ & ${\rm CSS}_{ N=4}$ & ${\rm CSS}_{ N=4}$ \\[.2em]
$e_2, e_3\rightarrow0$ & ${\rm CSS}_{ N=6}$ & ${\rm CSS}_{ N=6}$ & ${\rm CSS}_{ N=6}$ & ${\rm CSS}_{ N=6}$ \\
\hline
\end{tabular}
\end{center}
\caption{Contractions along the moduli space of the SO$^*(8)$ model. 
Note that CSO$^*(2,6)={\rm CSS}_{N=6}$, while ${\rm CSO}(2,0,6)={\rm CSS}_{N=0}$ only for $m_1=\ldots=m_4$.
All these contractions can be obtained by multiplying the embedding tensor by the moduli we then send to zero, except for the second line. 
In that case we multiply the embedding tensor by $e_1^{-1}$ and then take the limit $e_1^{-1}, e_2, e_3\rightarrow0$.
}\label{tab:contractions}
\end{table}

All these models share a $\rm \left[SU(1,1)/U(1)\right]^3$ factor, parametrized by $X_i,$ $E_i$, as a subsector of their moduli spaces. 
Hence we can consider the dependence on $X_i,$ $E_i$ acting \textit{after} the contraction, especially for what regards the mass spectra.
An important observation is that since $X_i$ and $E_i$ in the same  SU$(1,1)_i$ factor do not commute, if we perform a contraction along some $x_i$, say $x_1$ for concreteness, the dependence of the mass spectra on both $x_1$ and $e_1$ in the contracted model will be different from that of SO$^*(8)$, because $\Theta_{\rm contr}$ has a fixed degree of homogeneity with respect to $x_1$ (which $\Theta^{\mathfrak{so}^*(8)}$ does not have), and because the action of $E_1$ does not commute with the singular limit:
\begin{equation}
\begin{split}
\label{contraction to SO4 SO22}
X_1 \Theta_{\rm contr} &= x_1^{-1} \Theta_{\rm contr},\\[.3em]
E_1 \Theta_{\rm contr} &= E_1 \lim_{x_1\rightarrow0} x_1\, X_1 \Theta_0^{\mathfrak{so}^*(8)}\neq\lim_{x_1\rightarrow0} x_1\, X_1 E_1\Theta_0^{\mathfrak{so}^*(8)}
.\end{split}
\end{equation}
However, the dependence on $x_j$ and $e_j$, with $j\neq 1$, will be the same as the one observed in SO$^*(8)$, because their action commutes with the singular limit.
This fact has an important consequence: in the SO$^*(8)$ model the gravitino masses are controlled by the moduli $e_i$ up to an overall common factor, but once we perform a contraction along some $x_i$ direction, some of their ratios  are freezed in the contracted model.
In the example of \eqref{contraction to SO4 SO22}, corresponding to a contraction to  ${(\rm SO(4)\times SO(2,2))}\ltimes T^{16}$, the ratio $(M_2 M_3)/(M_1 M_4)$ depends only on $e_1$, and therefore it can be tuned in SO$^*(8)$ \textit{before} taking the singular limit, but there is no $e_i$ modulus governing its value in the contracted theory.
We can say that $(M_2 M_3)/(M_1 M_4)$ is a {\it modulus} of the SO$^*(8)$ theory, while it is only a {\it parameter} of the ${(\rm SO(4)\times SO(2,2))}\ltimes T^{16}$ model.
The most important example of this kind is the limit $x_1\sim x_2\sim x_3\equiv x \rightarrow 0$, which always reproduces the CSS models. 
In the CSS models the values of all four gravitino masses correspond to parameters of the theory that are not affected by the vevs of any modulus but for an overall common rescaling. 
However, as stated above, we can choose the masses $M_i$ in the SO$^*(8)$ theory by tuning (or taking a singular limit in) the $e_i$ moduli before we send $x\rightarrow 0$.

\begin{table}
	\renewcommand{\arraystretch}{1.3}\addtolength{\tabcolsep}{-1pt}%
\begin{center}
\rowcolors{1}{white}{gray!15}
\begin{tabular}{|c|c|c|c|c|}
	\hline
G$_{\rm gauge}$ & $\mathcal N$  & max. G$_{\rm res}$ & $r_a$ & ref.\\
\hline\hline
SO$^*(8)$ & 0 & $\rm SU(4)\times U(1)$  & - & \cite{DallAgata:2011aa}\\
CSO$^*(6,2)$ & 2 & SU$(3)\times \rm U(1)$  & $M_1=0$ & \cite{Hull:2002cv,Borghese:2012zs}\\
CSO$^*(4,4)$ & 4 & $\rm SU(2)\times U(1)$ & $M_1=M_2=0$& -\\
${(\rm SO(4)\times SO(2,2))}\ltimes T^{16}$ & 0 & $\rm SU(2)^2\times U(1)^2$  & $(M_2 M_3)/(M_1 M_4)$& \cite{DallAgata:2011aa}\\
$({\rm SO^*(4)\times U(1)})\ltimes{N^{20}}$ & 2 & $\rm SU(2)\times U(1)^2$   & $M_1=0$& -\\
${\rm U(1)^2}\ltimes N^{26}$ & 0 & U$(1)^2$ & $M_1/M_2,\ {M_3}/{M_4}$& \cite{DallAgata:2011aa}\\
${\rm U(1)^2}\ltimes N^{24}$ & 2 & U$(1)^2$  & $M_3/M_4$,\ $M_1=0$& -\\
${\rm U(1)^2}\ltimes N^{20}$ & 4 & U$(1)^2$  & $M_1=M_2=0$& -\\
CSS$_{\mathcal N}$&$\mathcal N$  & U(1) & $M_1/M_4$, $M_2/M_4$, $M_3/M_4$ & \cite{Cremmer:1979uq,Andrianopoli:2002mf} \\[2pt]\hline
\end{tabular}
\caption{Maximal residual gauge symmetries, supersymmetries and fixed gravitino mass ratio parameters $r_a$ in the models that can be reached by contraction of SO$^*(8)$ along SU$(1,1)^3/{\rm U}(1)^3$.
We also give reference to where the corresponding vacua have been discussed.
The indicated $\rm G_{res}$ do not include additional trivial U(1) factors that may appear in correspondence with ungauged vectors.}\label{tab:fixed parameters in contracted models}
\end{center}
\end{table}

In general, we can say that the contracted models have at least six moduli $x_i,\,e_i$, and as many gravitino mass ratio parameters $r_a$ as the (maximum) number of contractions in the $x_i$ directions that must be done to reach the model from SO$^*(8)$. 
The corresponding $x_a$ moduli are overall rescalings of the contracted embedding tensor. 
Note also that while moduli are associated with E$_{7(7)}$ transformations and cannot affect the gauge group structure constants, parameters can affect them, as we have seen for the CSS models in Section~\ref{sec:css_gaugings}. 
In fact, in some models the dimensions of the nilpotent algebras reduce for specific values of the parameters.
The situation is summarized in Table \ref{tab:fixed parameters in contracted models}.

When supersymmetry gets enhanced, also the moduli space gets promoted to a scalar manifold that is consistent with the corresponding supersymmetries.
For instance, the $N=4$ vacua of Table~\ref{tab:fixed parameters in contracted models} have 14 massless fields parameterizing the moduli space
\begin{equation}
	\frac{\rm{SU}(1,1)}{\rm{U}(1)} \times \frac{\rm{SO}(6,2)}{\rm{SO}(6) \times \rm{SO}(2)}.
\end{equation}

Mass spectra for all the contracted models can be obtained from the mass formula (\ref{U1spectrum})--(\ref{U1metric}) by using (\ref{mu1})--(\ref{so62charges}), where appropriate limits have been taken.
For instance, the (SO(4) $\times$ SO(2,2)) $\ltimes T^{16}$ model has a mass spectrum that follows by multiplying (\ref{mu1})--(\ref{mu4}) by $x_1^2$ and then taking the limit $x_1 \to 0$, so that in this case
\begin{eqnarray} 
\mu_1^2 &=& \frac{1+x_2^2x_3^2}{8}, \\[2mm]
\mu_2^2 &=& \frac{x_2^2-x_3^2}{8\, x_3^2}, \\[2mm]
\mu_3^2 &=& \frac{x_3^2-x_2^2}{8\, x_2^2}, \\[2mm]
\mu_4^2 &=& \frac{1+x_2^2x_3^2}{8\, x_2^2 x_3^2}.
\end{eqnarray}
Moreover, since the action of $E_1$ does not commute with the contraction along $x_1$, we need to substitute $e_1$ in (\ref{so62charges}) with a fixed parameter: $e_1 \to r_1 \equiv (M_2 M_3)/(M_1 M_4)$.
The fields $x_1$ and $e_1$ are still moduli and, using (\ref{contraction to SO4 SO22}), we can show that they appear in the mass spectrum as an overall factor $\sqrt{\left(1+e_1^2\right)/\left(2 \, e_1^2 \, x_1^2\right)}$.
The spectrum of the [SO$^*(4) \times $U(1)] $\ltimes N^{20}$ model then follows by further multiplying the $\vec q$ charges by $1/r_1$ and then taking the limit $r_1^{-1}, e_2, e_3\rightarrow0$, so that 
\begin{eqnarray}
\vec{q}_1 = - \vec{q}_2 = (0,0,0,0) \, , 
& &                      
\vec{q}_3 = - \vec{q}_4 = k\,(+1,+1,-1,-1) \, , 
\nonumber \\
\vec{q}_5 = - \vec{q}_6 = \frac1k\,(+1,-1,+1,-1) \, , 
& & 
\vec{q}_7 = - \vec{q}_8 = \tilde k\, (+1,-1,-1,+1) \, ,
\end{eqnarray}
where 
\begin{equation}
	k = \lim_{e_2,e_3 \rightarrow 0} \frac{e_3}{e_2} \quad {\rm and} \quad \tilde k = \lim_{1/r_1,e_2,e_3 \rightarrow 0} \frac{1}{r_1^2\, e_2\, e_3}
\end{equation}
are parameterized by finite $E_2$ and $E_3$ transformations.
It is straightforward to see that in this limit one of the gravitino masses vanishes, $M_1 = 0$,  therefore the vacuum preserves ${\cal N}=2$ supersymmetry.
Note that the mass formula for this model requires charges with respect to U(1)$^4$, though the diagonal U(1) is now a global symmetry.


\subsection{Families of CSO$^*$ gaugings} 
\label{sub:families_of_cso_gaugings}

The findings of this analysis of the CSO$^*$ models can be summarized in the following way: the CSO$^*$ gaugings provide a family of models admitting Minkowski vacua that have gravitino masses that can be arbitrarily tuned by moving in moduli space.
In particular, by moving in moduli space we can reproduce the mass spectra (of all fields) of any CSS model, which we actually recover as special limits towards the boundary.
The rank of the CSO$^*$ gauge group and of the non-Abelian residual gauge symmetry group is directly related to the number $N=2n$ of non-zero gravitino masses and hence to the different number of supersymmetries preserved on the Minkowski vacuum:
\begin{equation}
	\rm{G_{gauge}} = {\rm CSO}^*(8-2n,2n) \longrightarrow \rm{G_{res}} = {\rm U}(1) \times {\rm SU}(2n).
\end{equation}
Just like the CSS models, we can actually produce these models by superpositions of the embedding tensor defining the most symmetric one.

Since CSO$^*(2,6) \simeq {\rm CSS}_{{\cal N} = 6}$, we have to be more precise on the superposition details.
As we saw in section \ref{sec:css_gaugings}, if we take two CSS$_{{\cal N} = 6}$ models specified by different matrices $M$ in the same electric E$_6$ frame and superimpose them, we obtain another CSS model with lower supersymmetry.
Each CSS$_{{\cal N} = 6}$ model is specified by a matrix $M_\mu{}^\nu$, which can be taken to be of the form
\begin{equation}
	M = \left(\begin{array}{ccc}
	 & m \,{\mathbb 1}_{6} &  \\
	 -m\,{\mathbb 1}_{6} &  & \\
	  &  & {\mathbb 0}_{15}
	\end{array}\right).
\end{equation}
The corresponding gauge generators have a non-trivial action on the vectors in the $(\mathbf{2},\mathbf{6})_{-1}$ and the rotation to the SU$^*(8)$ frame (\ref{rotation to SUstar}) mixes them with those in the $(\mathbf{2},\mathbf{6}')_{+1}$.
The result is a model that is equivalent to a CSO$^*(2,6)$, with electric action in the SU$^*(8)$ frame, by means of a SU(8) duality transformation.
If we sum the embedding tensor of such model with that of a different CSO$^*(2,6)$, which has a non-trivial action on a different set of vectors (some of which may overlap with the previous ones), we get a new embedding tensor that is not equivalent to the one obtained by the superposition of CSS models in their electric frame.
In fact, we need two different SU(8) rotations, related to the two different tensors that have been superposed and therefore the two duality transformations are not compatible with each other.
However, remarkably, also in this case both the critical point condition as well as the fact that the potential vanishes at such critical points are identically satisfied for these superpositions.


\subsection{Additional SU*(8) models} 
\label{sub:extra_sustar8_models}

So far, we discussed electric gaugings of SU$^*(8)$, namely [SO$^*(8)\simeq{\rm SO}(6,2)]_{c=1}$ and the CSO$^*(2p,2r)$ models, together with their contractions along the moduli space. 
We will now show that there are other models with Minkowski vacua that can be constructed by dyonic superpositions of the previous ones.

Group theoretically, the SO$(6,2)_{c=1}$ and CSO$^*(2p,2r)$ models are defined by an embedding tensor sitting in the $\overline{\bf36}$ of SU$^*(8)$ and the quadratic constraint is trivially satisfied.
Analogously to the `$\theta\xi=0$' gaugings defined in \cite{DallAgata:2011aa} using the embedding tensor components in the $\mathbf{36}$ and $\mathbf{36}'$ of SL$(8,\mathbb R)$, we can expect that there exist other models where G$_{\rm gauge}\subset{\rm SU}^*(8)$, but where the gauge connection contains both electric and magnetic fields.
Following the analogy, we expect to obtain ``superpositions'' of two CSO$^*(2p,2r)$ groups where the two SO$^*(2p)$ factors commute, while the translations combine to form a larger nilpotent algebra.
The resulting models have a gauge group of the form
\begin{equation}
\left[{\rm SO}^*(2p)\times{\rm SO}^*(2p')\right]\ltimes N^{4[(4-r)r+(4-r')(r+r'-4)]},
\qquad p+p'\le4.
\end{equation}
Always following the analogy with \cite{DallAgata:2011aa}, we expect the quadratic constraint to be satisfied whenever the two semisimple (or U(1)) factors are gauged by the electric vector fields
$A_\mu^{[ab]},\ a,b=1.\ldots2p$ and the magnetic $A_{\mu\ [a',b']},\ a',b'=2p+1,\ldots2(p+p')$ respectively, where we decomposed indices in the $\bf 8$ of $\rm SU^*(8)$ as $A=(a,a',\ldots)$.
We explicitly built these models and checked their consistency. 
As usual, we can classify them according to the gauge group:
\begin{equation}
{\rm U}(1)^2\ltimes N^{20},\quad
[{\rm SO}^*(4)\times{\rm U}(1)]\ltimes N^{20},\quad
{\rm SO}^*(4)^2\ltimes T^{16},\quad
[{\rm SO}^*(6)\times{\rm U}(1)]\ltimes T^{12}.
\end{equation}
All of them have a Minkowski vacuum and a $[{\rm SU(1,1)/U(1)}]^3$ moduli space. In fact, the first two models already appear in Table \ref{tab:contractions} and have $N=4$ and $N=2$ supersymmetry respectively.
Moreover, we can see that
\begin{equation}
{\rm SO}^*(4)^2\ltimes T^{16} 
\ \simeq\  [{\rm SU(2)^2\times SL(2)^2}]\ltimes T^{16}
\ \simeq\  [{\rm SO(4)\times SO(2,2)}]\ltimes T^{16} \, ,
\end{equation}
hence also this model already appeared in Table \ref{tab:contractions}.

The ${[\rm SO^*(6)\times U(1)]}\ltimes T^{16}$ model is genuinely new.
Apparently the only way to reach it from ${\rm SO^*(8)}$ is by slightly perturbing the $c$-parameter, namely setting $c=1+\epsilon$, and then contracting $\epsilon\sim e_1^{-1}\sim e_2\sim e_3\to0$. However, modifying the $c$ parameter breaks the vacuum condition, and this model cannot be seen as arising from the moduli space of SO$^*(8)$.
Even if this is the case, this model fits well with the class of theories discussed so far. The Minkowski vacuum of the theory has fully broken supersymmetry and residual gauge group 
\begin{equation}
	{\rm G_{res} = SU(3)\times U(1) \times U(1)}
\end{equation}
at the maximally degenerate point, which further breaks to U$(1)^4$ along its $\rm [SU(1,1)/U(1)]^3$ moduli space.

Going to the boundary of the moduli space of this model always brings us back to the theories of Table \ref{tab:contractions}, with one notable exception, given by an $N=0$ model with gauge group ${[\rm SO^*(4)\times U(1)]}\ltimes N^{20}$. 
This is obtained by contracting along \emph{one} of the $x_i$ moduli of the [SO$^*(6) \times $U(1)$]\ltimes T^{12}$ model above.



\section{The [SO(3,1) $\times$ SO(1,3)] $\ltimes T^{16}$ model} 
\label{sec:the_so_3_1_times_so_1_3_ltimes_t_16_model}

While most of the $N=8$ models with Minkowski vacua are in the CSO$^*$ family or in its contractions, we have an outstanding example that does not fall in this class.
This model, with gauge group [SO(3,1) $\times$ SO(1,3)] $\ltimes T^{16}$, has been first identified in \cite{Dall'Agata:2012sx} as a contraction of the SO(4,4)$_c$ theories.
Its construction follows lines similar to those of the contraction procedure of section~\ref{sec:gaugings_and_contractions}, as we now review.

The starting point is the SO(4,4)$_c$ family, where the embedding tensor in the SL($8,{\mathbb R}$) frame is determined by
\begin{equation}\label{thxi44}
	\theta = {\rm diag}\{1,1,1,1,-1,-1,-1,-1\} \quad {\rm and} \quad \xi = c\ \theta^{-1}.
\end{equation}
As shown in \cite{Dall'Agata:2012sx}, \cite{Borghese:2013dja}, these models have 3 de Sitter vacua in the range $c\in [0,\sqrt2 -1[$,
one already known for $c=0$ and two genuinely new.
When we reach $c = \sqrt2 - 1$ the new vacua disappear, but defining an appropriate limit for $c \to \sqrt2 - 1$ they become Minkowski vacua of a contracted model.
Starting from $\Theta^{\mathfrak{so}(4,4)_c}$, defined as in $(\ref{thxi44})$, we can introduce a new embedding tensor depending on two parameters $\tau$ and $x$, related to the E$_{7(7)}$ generators that preserve SO(3) $\times$ SO(3) $\subset$ SO(4,4) and survive the D$_4$ projection generated by \cite{Dall'Agata:2012sx}
\begin{eqnarray}
	Z_1 &=& \sigma_1 \otimes ({\mathbb 1}_3\oplus -1), \\[2mm]
	Z_2 &=& \sigma_3 \otimes ({\mathbb 1}_3\oplus -1),
\end{eqnarray}
acting on the SL($8,{\mathbb R}$) indices.
These generators are \cite{Dall'Agata:2012sx}
\begin{eqnarray}
	g_5 &=& t_1{}^1+t_2{}^2+t_3{}^3+t_5{}^5+t_6{}^6+t_7{}^7-3(t_4{}^4+t_8{}^8),  \\[2mm]
	g_6 &=& t^{1238}+t^{4567}.
\end{eqnarray}
We then act with
\begin{equation}
	G(x,\tau) = \exp\left(\frac{3}{\sqrt2}\,g_5\, \log x +\sqrt{6} \,g_6\, \log \tau\right)
\end{equation}
on $\Theta^{\mathfrak{so}(4,4)_c}$, as described in (\ref{Thetax}) and name the resulting embedding tensor $\Theta^{\mathfrak{so}(4,4)_c}(x,\tau)$.
This is the embedding tensor that generates the scalar potential discussed in \cite{Dall'Agata:2012sx}, whose vacua appear at $x=\tau=1$ and at specific points in the range
\begin{equation}
	x \in \left[1+\frac{2}{\sqrt3},3+2\sqrt2\right[, \quad \tau \in \left]0,\sqrt{5-2\sqrt6}\right],  \quad {\rm or} \quad \tau \in \left[\sqrt2 + \sqrt3, +\infty\right[,
\end{equation} 
depending on the value of $c$, as follows from
\begin{eqnarray}
\label{eqx}	&&1 - 3 (c^2 -2) x + 3(2c^2 -1) x^2 + c^2 x^3 = 0, \\[3mm]
\label{eqtau}	&&\tau = \frac{-3\pm 2\sqrt2 + 2 x - (3 \pm 2 \sqrt2) x^2}{x^2-6x+1}.
\end{eqnarray}
If we use (\ref{eqx})--(\ref{eqtau}) in $\Theta^{\mathfrak{so}(4,4)_c}(x,\tau)$, we get that the corresponding critical point moved at the origin of the moduli space.
As $c \to \sqrt2 -1$ we have that $x \to 3 + 2 \sqrt2$, $\tau \to 0,+\infty$ and that the corresponding vacua disappear from the allowed region of the moduli space.
From the embedding tensor point of view, $\Theta^{\mathfrak{so}(4,4)_c(x(c),\tau(c))}$ contains divergent terms.
We then introduce the contracted tensor
\begin{equation}
	g' \ \Theta_{\rm contr} = \lim_{c\to\sqrt2-1}\, g \  \Theta^{\mathfrak{so}(4,4)_c}\left[x(c),\tau(c)\right],
\end{equation}
where $g = g' \tau$.
The result is a new embedding tensor that gauges [SO(3,1) $\times$ SO(1,3)] $\ltimes T^{16}$, which, in fact, is a contraction of SO(4,4).

Using the methods of \cite{DallAgata:2011aa}, we then see that this model has a Minkowski vacuum, whose residual symmetry group is
\begin{equation}
	\rm{G_{res}} = {\rm SO}(3) \times {\rm SO}(3).
\end{equation}
This is the first instance of a Minkowski vacuum whose residual gauge group does not have any Abelian factor.
The spectrum at this point is summarized in Table~\ref{tab_massesso3113}.

\begin{table}
	\renewcommand{\arraystretch}{1.6}\addtolength{\tabcolsep}{1pt}%
\begin{center}
\rowcolors{1}{white}{gray!15}
\begin{tabular}{|c|c|l|}
\hline
	spin		& $[m^2]_{\text{SO(3)}\times \text{SO(3) irreps.}}$ \\
\hline\hline
	$2$ & $[0]_{(\mathbf{1},\mathbf{1})}$\\
 	$\frac32$ & $\left[\frac{1}{2}\right]_{(\mathbf{2},\mathbf{2})+(\mathbf{2},\mathbf{2})}$ \\
	$1$ & $[0]_{(\mathbf{3},\mathbf{1})+(\mathbf{1},\mathbf{3})}$, \ $\left[\frac43\right]_{(\mathbf{3},\mathbf{1})+(\mathbf{1},\mathbf{3})}$, \ $[1]_{(\mathbf{3},\mathbf{1})+(\mathbf{1},\mathbf{3})+(\mathbf{3},\mathbf{3})+(\mathbf{1},\mathbf{1})}$ \ \\
	$\frac12$ & $[0]_{(\mathbf{2},\mathbf{2})+(\mathbf{2},\mathbf{2})},\  \left[\frac12\right]_{2 \times (\mathbf{2},\mathbf{2}) + (\mathbf{4},\mathbf{2})+(\mathbf{2},\mathbf{4})}, \ \left[\frac{11}{6}\right]_{2 \times (\mathbf{2},\mathbf{2}) + (\mathbf{4},\mathbf{2})+(\mathbf{2},\mathbf{4})}$\\
	$0$ & $[0]_{2 \times [(\mathbf{3},\mathbf{1})+(\mathbf{1},\mathbf{3})]+ 2 \times (\mathbf{3}, \mathbf{3})+3 \times (\mathbf{1},\mathbf{1})}$, \ $[1]_{(\mathbf{3},\mathbf{3})+ (\mathbf{3},\mathbf{1})+ (\mathbf{1},\mathbf{3}) + (\mathbf{1},\mathbf{1})}$, \ $\left[\frac43\right]_{(\textbf{5},\textbf{1})+(\mathbf{1},\mathbf{5})+2\times (\mathbf{1},\mathbf{1})}$, \ $\left[\frac83\right]_{(\mathbf{3},\mathbf{3})}$
	\\\hline
\end{tabular}
\end{center}
\caption{Spectrum of the  [SO(3,1) $\times$ SO(1,3)] $\ltimes T^{16}$ model at the SO(3) $\times$ SO(3) Minkowski critical point.}
\label{tab_massesso3113}
\end{table}

The 33 massless scalars can be split in 22 Goldstone bosons of the broken non-compact gauge symmetries and 11 real moduli fields.
Two of these 11 moduli correspond to the $x$ and $\tau$ fields, which, as usual, remain massless after the contraction and appear as overall rescalings of the masses.
The remaining 9 fields are in the $(\mathbf{3},\mathbf{3})$ representation of SO(3) $\times$ SO(3) and their expectation values break further this residual symmetry to U(1) $\times$ U(1).
If we label the fundamental representation of the two factors by $i=1,2,3$ and $a=1,2,3$, the 9 moduli in the $(\mathbf{3},\mathbf{3})$ can be described by a field $\Phi_{ia}$.
Given any expectation value 
\begin{equation}
	\langle\Phi_{ia}\rangle \neq 0,
\end{equation}
we can always use the SO(3) $\times$ SO(3) gauge symmetry to rotate it to a definite direction, for instance $i=1$ and $a=1$.
The resulting vacuum is then invariant under U(1) $\times$ U(1) rotations on the $i=2,3$ and $a=2,3$ planes.
This symmetry breaking mechanism introduces additional factors in the masses, depending on the vev of the field $\Phi_{ia}$.
We therefore chose one of the generators $t$ corresponding to the $\Phi_{ia}$ scalars and constructed the group geodesic $G(\phi) = \exp(t \log \phi)$, by which we act on $\Theta_{\rm contr}$, obtaining a new embedding tensor from which we extract the mass spectrum.
Unfortunately, the mass spectrum does not simply fit the mass formula (\ref{U1spectrum}), in contrast with the examples of \cite{Dall'Agata:2012cp}.
In fact, by inspecting (\ref{U1spectrum}) we can see that whenever all the gravitino masses take the same value $m_{3/2}^2$, the masses of the spin 1 fields must satisfy the following relation:
\begin{equation}
\#({\rm massless\ vectors}) = 4 + \#({\rm vectors\ with\ } m^2=4m_{3/2}^2).
\end{equation}
This is clearly not the case for the mass spectrum of Table \ref{tab_massesso3113}.
However, if we remove the value of the mass of the same fields at the $\phi =1$ point, the shifts in the masses are once again described by a relation such as (\ref{U1spectrum}):
\begin{equation}
	m^2= \frac1\tau \frac{1+ x^4}{2 x^2}\left(m_{SO(3)^2}^2 + q_L^2 m_L^2 + q_R^2 m_R^2\right),
\end{equation}
where $m_{SO(3)^2}^2$ is the value of the same field as in Table~\ref{tab_massesso3113},
\begin{equation}
	\label{mLmR}
	m_L^2 = \frac16\, \frac{1-\phi^2}{2}, \qquad m_R^2 = \frac16(\phi^2-1) \, ,
\end{equation}
and $q_{L,R}$ are the charges with respect to the diagonal and antidiagonal combination of the surviving U(1) $\times$ U(1) $\subset$ SO(3) $\times$ SO(3).
The complete spectrum of charges for the various fields of this model is given in Table~\ref{tab_massescomplete}.

\begin{table}
	\renewcommand{\arraystretch}{1.3}\addtolength{\tabcolsep}{-1pt}%
\begin{center}
\rowcolors{1}{white}{gray!15}
\begin{tabular}{|c|c|c|}
\hline
	spin	& $m_{SO(3)^2}^2$& \small{U(1)$^2$} charges	\\
\hline\hline
	$2$ & 0 & {\small $(0,0)$} \\
 	$\frac32$ & $\frac12 $ & {\small  $2 \times [(\pm1,0)+(0,\pm1)]$ }\\
	$1$ & 0 & {\small $2 \times (0,0) + (\pm1,\pm1)$ }\\
	$1$ & $\frac43$ &{\small $2 \times (0,0) + (\pm1,\pm1)$ }\\
	$1$ & $1$ & {\small $4 \times (0,0) + (\pm2,0) + (0,\pm2) + 2 \times (\pm1,\pm1)$ }\\
	$\frac12$ & 0&{\small $2\times[(\pm1,0)+(0,\pm1)]$} [Goldstini]\\
	$\frac12$ & $\frac12$& {\small $4\times[(\pm1,0)+(0,\pm1)] + (\pm1,\pm2)+(\pm2,\pm1)$ }\\
	$\frac12$ & $\frac{11}{6}$&{\small $4\times[(\pm1,0)+(0,\pm1)]+ (\pm1,\pm2)+(\pm2,\pm1)$ }\\
	$0$ & $0$ & {\small $6 \times (0,0) + 4 \times (\pm1,\pm1) + (\pm2,0)+(0,\pm2)$} [Goldstones]\\
	$0$ & $0$ &{\small $3 \times (0,0) + (\pm2,0)+(0,\pm2)$}\\
	$0$ & $1$ & {\small $4 \times (0,0) + (\pm2,0)+(0,\pm2)+ 2 \times (\pm1,\pm1)$}\\
	$0$ & $\frac43$ &{\small $4 \times (0,0) + (\pm1,\pm1)+(\pm2,\pm2)$}\\
	$0$ & $\frac83$ & {\small $(0,0) + (\pm1,\pm1)+(\pm2,0)+(0,\pm2)$}
	\\\hline
\end{tabular}
\end{center}
\caption{Values of the charges of the fields at the generic U(1) $\times$ U(1) invariant vacuum.}
\label{tab_massescomplete}
\end{table}

The mass spectrum we recover in this way reveals that unfortunately the vacuum we are discussing is only marginally stable at $\phi =1 $ and becomes unstable for any $\phi \neq 1$.
Two of the massless modes at $\phi =1$ in fact have charges $(\pm2,0)$ and therefore acquire masses proportional to $m_L^2$, while two others have charges $(0,\pm2)$ and therefore acquire masses proportional to $m_R^2$.
As it is clear from (\ref{mLmR}), the first two are negative for $\phi>1$ and the second two are negative for $\phi<1$.
Nevertheless, as expected, the spectrum fulfills 
\begin{equation}
	{\rm Str}\, {\cal M}^{2k} = 0
\end{equation}
for $k=0,1,2,3$ at any point in the moduli space. At $\phi =1$, we find once more ${\rm Str}\, {\cal M}^8 > 0
$ and the 1-loop correction of the scalar potential turns out to be negative.

Finally, contracting this model along $\phi$ brings us back to the CSS $N=4$ model with the two non-vanishing mass parameters equal to each other.


\section{Summary and discussion} 
\label{sec:discussion}

In this work we analyzed the moduli space of spontaneously broken $N=8$ supergravity theories in 4 dimensions with classical Minkowski vacua.
We showed that all known classical vacua can be connected by sending some of the moduli to their boundary value.
In particular, we found that most of the models arise as contractions of the SO$^*(8)$ gauging, whose electric frame has been rotated with respect to the standard SL$(8,{\mathbb R})$ one by using the U(1) rotation introduced in \cite{Dall'Agata:2012bb} with $c=1$.
Among others, these contractions include all the CSS models and provide a more general supersymmetry breaking scheme, which allows for residual non-Abelian gauge groups.

While most of the models found so far obey a general mass formula that relates the masses of all fields to their U(1) charges, as previously argued in \cite{Dall'Agata:2012cp}, we also found the first instance of a Minkowski vacuum that does not fit in this simple framework.
Its residual gauge group does not have Abelian factors and the masses do not obey any simple  generalization of the mass formula presented in \cite{Dall'Agata:2012cp}.

As explained in the main text, in the wide class of old and new gauged $N=8$ supergravities considered in this paper, admitting classical Minkowski vacua with fully broken supersymmetry, the classical moduli space is lifted by 1-loop corrections, which give moduli-dependent negative contributions to the effective potential. 
Obviously, this cannot provide any highly desired example of 1-loop locally stable de~Sitter or Minkowski vacuum.
We could ask, however, whether the resulting 1-loop effective potential has locally stable $AdS$ critical points and, if so, what are the corresponding masses of the classical moduli.
A preliminary analysis for the SO$^*(8)$ model shows that this is in fact the case, but also that some of the moduli acquire a mass that is negative and below the Breitenlohner--Freedman bound, so that the resulting 1-loop critical point is unstable.
More generally, it would be interesting to understand, going beyond the models constructed so far and considered in this paper, whether and why all Minkowski vacua with fully broken $N=8$ supersymmetry are indeed destabilized by quantum corrections.
We cannot exclude the existence of other models, with different mass patterns, which may allow for positive or vanishing 1-loop contributions to the scalar potential in some region of their classical moduli space.
However, as noted in \cite{Dall'Agata:2012cp}, there seems to be a strong correlation between the sign of the 1-loop potential and the sign of the first non-vanishing supertrace.
To understand whether the destabilization phenomenon we observed here has a universal character a possible strategy could then be to look for new models with Str ${\cal M}^8 < 0$.

We must also note that, so far, we have been able to produce partial supersymmetry breaking on Minkowski vacua only in steps of 2, i.e.~we can break $N=8$ supersymmetry to $N=6, 4, 2$ or 0, but we do not have any example with $N=3$ or $N=1$ residual supersymmetry.
While this could be explained for most of the known models as the consequence of the Abelian factors in the residual gauge group, forcing the gravitino masses to be of Dirac type, we cannot exclude a priori that there may be models allowing for such a supersymmetry breaking pattern.
In particular, the model of section~\ref{sec:the_so_3_1_times_so_1_3_ltimes_t_16_model} is the first model where we have a massive vector that is a singlet with respect to the full residual gauge group, thus we may now expect to find also models with massive gravitinos that are singlets.
If there is a universal relation between the mass spectrum and the charges of the various fields with respect to the residual gauge group, these models may allow for partial breaking to $N=1$ or $N=3$.

Finally, we do not know the higher-dimensional origin of most of the models presented here.
This would be particularly interesting, given the fact that if we can interpret the classical 4-dimensional moduli fields as parameters of some internal manifold, their motion towards the boundary could be interpreted as special singular limits on the internal manifold properties.
This may explain the connectedness of different moduli spaces as geometric transitions in higher dimensions.



\bigskip
\section*{Acknowledgments}

\noindent 
We would like to thank Mario Trigiante for discussions.
This work is supported in part by the ERC Advanced Grant no. 226455, \textit{``Supersymmetry, Quantum Gravity and Gauge Fields''} (\textit{SUPERFIELDS}), by the ERC Advanced Grant \emph{Electroweak Symmetry Breaking, Flavour and Dark Matter: One Solution for Three Mysteries} (\emph{DaMeSyFla}), by the European Programme UNILHC (contract PITN-GA-2009-237920), by the Padova University Project CPDA105015/10 and by the MIUR grant RBFR10QS5J \emph{``String Theory and Fundamental Interactions''}.
The research of F.~C.~is supported by the Fond.~Angelo Della Riccia.


\appendix

\section{CSS models from M-theory} 
\label{sub:m_theory_lift}

In \cite{Scherk:1979zr}, \cite{Cremmer:1979uq}, \cite{Dall'Agata:2005ff}, \cite{D'Auria:2005dd}, \cite{D'Auria:2005rv} it has been established that the CSS gaugings with up to 3 independent mass parameters can arise from reductions of M-theory on a twisted 7-dimensional torus, with appropriate fluxes turned on.
However, the generic CSS model contains 4 parameters and therefore we still miss an explanation for one of them.
In the following we are going to provide a possible origin for the missing parameter in the same setup.

Our starting point is the assumption that the 7-dimensional manifold on which we compactify M-theory is a deformation of the torus, possibly also by means of non-geometric fluxes \cite{Dall'Agata:2007sr}.
For this reason, we classify the 912 possible gauging parameters of the embedding tensor with geometric and non-geometric fluxes, according to their GL($7,{\mathbb R}$) representations, because GL($7,{\mathbb R})$ is the group of change of coordinates of the internal manifold.
Following \cite{Dall'Agata:2007sr}, the branching of the representation \textbf{912} of E$_{7(7)}$ under SL$(7,{\mathbb R}) \times$ O(1,1) gives
\begin{eqnarray}
	{\bf 912}&\rightarrow &{\bf 1}_{-7}+{\bf 1}_{+7}+{\bf 35}_{-5}+ {{\bf 35}}^\prime_{+5}+ ( {{\bf 140}}^\prime+ {{\bf 7}}^\prime)_{-3}+ ({\bf 140}+{\bf 7})_{+3}+{\bf 21}_{-1}+ {{\bf 21}}^\prime_{+1}+\nonumber\\&&{\bf 28}_{-1}+ {{\bf 28}}^\prime_{+1}+{\bf 224}_{-1}+ {{\bf 224}}^\prime_{+1}\label{branch912}\,.
\end{eqnarray}
Each representation has a corresponding tensor representation, which is interpreted as one of the geometric or non-geometric fluxes:
\begin{equation}\label{fluxtable}
	\begin{array}{cccccccc}
	\mathbf{1}_{\mathbf{+7}} & g_7, & \phantom{PIP} & (\mathbf{140}+\mathbf{7})_{+\mathbf{3}} & \tau_{jk}^i+ \delta^i_j\, \tau_k, &  \phantom{PIP} & \mathbf{28}_{-\mathbf{1}} & \theta_{(ij)},\\[3mm]
	\mathbf{1}_{\mathbf{-7}} & g_7, && (\overline{\mathbf{140}}+\overline{\mathbf{7}})_{-\mathbf{3}} & Q^{jk}_i+ \delta_i^j\, Q^k,  && \overline{\mathbf{28}}_{+1} & \xi^{(ij)},\\[3mm]
	\bf{35}_{-5} & h^{ijkl}, && {\bf{224}}_{-1} & f^{i}_{jkl}, && \bf{21}_{-1} & \theta_{[ij]}, \\[3mm]
	\overline{\bf{35}}_{+5} & g_{ijkl}, && \overline{\bf{224}}_{+1} & R_{i}^{jkl}, && \overline{\bf{21}}_{+1} & \xi^{[ij]}.
	\end{array}
\end{equation}
We should stress that what is considered to be geometric or not depends on the framework.
For instance, the parameters $\theta_{ij}$ have a geometric interpretation if we consider a reduction on the 7-sphere, while they are clearly non-geometric if interpreted as a deformation of the 7-torus.

The identification of the fluxes that give rise to the four CSS mass parameters can be established by comparing the scalar potential obtained by reducing M-theory on a twisted ${\mathbb T}^7$ as a function of the fluxes above and the scalar potential of the CSS models as a function of the mass parameters $m_i$, $i=1,2,3,4$.
The final outcome depends on the choice of duality frame and also on the way we decide to fix the residual gauge symmetry.
We find that the simplest and most geometric match gives
\begin{eqnarray}
	\label{tm1}
	\tilde m_1 &=& \tau_{73}{}^6 = - \tau_{76}{}^3, \\
	\tilde m_2 &=& \tau_{71}{}^4 = - \tau_{74}{}^1, \\
	\tilde m_3 &=& \tau_{75}{}^2 = - \tau_{72}{}^5, \\
	\tilde m_4 &=& g_7 = - \theta_{77}, \label{tm4}
\end{eqnarray}
where
\begin{eqnarray}
	\tilde m_1 &=& m_1 - m_2 - m_3 + m_4, \\
	\tilde m_2 &=& m_1 - m_2 + m_3 - m_4, \\
	\tilde m_3 &=& m_1 + m_2 - m_3 - m_4, \\
	\tilde m_4 &=& m_1 + m_2 + m_3 + m_4.
\end{eqnarray}
As it is clear from these identifications, the fourth parameter is associated to a non-geometric flux: $\theta_{77}$.
By using the U-duality group we can construct an orbit of equivalent gaugings, where the identifications will change.
However, we did not find a frame where all fluxes had a geometric interpretation and actually, (\ref{tm1})--(\ref{tm4}) provide the simplest setup.

The non-geometric flux $\theta_{77}$ does not have a direct interpretation in M-theory. However, a possible explanation for its origin may come from extending the ideas in \cite{Gibbons:2001wy}, where a possible M-theory origin for CSO($p,q,r$) gaugings was found. 
In the SL($8,{\mathbb R}$) frame, both $\theta_{77}$ and $g_7$ are related to the same $\theta_{AB}$ tensor, $A,B=1,\ldots,8$, specifying the CSO gauge groups (where $\theta_{88} = -g_7$).
In particular, the CSO(2,0,6) gauging is precisely determined by a tensor $\theta_{AB}$ with 2 non-zero positive values on the diagonal and the rest of the entries vanishing.
This corresponds to the CSS model with all mass parameters identified, $m_1 = m_2 = m_3= m_4$, i.e.~where only $\tilde m_4$ is non-zero.
Following \cite{Gibbons:2001wy}, the correct gauging should be reproduced by reducing M-theory on a 7-dimensional manifold embedded in ${\mathbb R}^8$ via
\begin{equation}
	z^A z^B \, \theta_{AB} = R^2.
\end{equation}
Unfortunately, for the CSO(2,0,6) gauging the resulting manifold is simply $S^1 \times {\mathbb R^6}$ and therefore we cannot expect the result to hold.
However, just as, when we compactify M-theory on $S^7 = {\rm SO}(8)/{\rm SO}(7)$, we have to use the full coset structure of the internal manifold to produce a consistent reduction to 4-dimensions, we could  therefore try to interpret the reduction procedure on a coset obtained by the quotient of the CSO(2,0,6) $ = U(1) \ltimes T^{12}$ group with an appropriate subgroup.
An alternative path could be provided by the use of the extended double geometry presented in \cite{Dall'Agata:2007sr,Gibbons:2001wy,Coimbra:2012af,Aldazabal:2013mya,Berman:2013uda}.
Since this goes beyond the scope of the present paper, we leave its analysis to future investigations.


\section{$N=1$ truncations} 
\label{sub:_n_1_truncations}

A useful tool to perform partial but quick checks of the results of the present paper is the existence of an $N=1$ truncation that keeps only 7 complex scalar fields.
For this reason, we present here the superpotential coming from general flux compactifications of M-theory on a $G_2$ manifold that is a twisted version of a ${\mathbb T}^7$, following the construction in \cite{Dall'Agata:2005fm}.
The $G_2$ manifold comes from the quotient of ${\mathbb T}^7$ by
\begin{equation}
\begin{array}{rcl}
{\mathbb Z}_2 (y^I) &=& \{-y^5,-y^6,-y^7,-y^8,c+y^9,y^{10},y^{11}\},\\[3mm]
{\mathbb Z}^\prime_2 (y^I) &=& 
\{-y^5,-y^6,y^7,y^8,c-y^9,c-y^{10},c + y^{11}\},\\[3mm]
{\mathbb Z}^{\prime\prime}_2 (y^I) &=& 
\{-y^5,y^6,c-y^7,y^8,-y^9,c+y^{10},-y^{11}\},
\end{array}
\label{orbifold}
\end{equation}
where $c$ can be either $0$ (singular variety) or $1/2$ (smooth manifold) and in the following we take $c=0$, for simplicity.

Obviously not all the fluxes presented in (\ref{fluxtable}) survive the (\ref{orbifold}) identifications and at the same time 63 out of the 70 scalar fields are projected out.
In detail, only seven 3-cycles of ${\mathbb T}^7$ survive, and the seven surviving scalar fields can be associated to such cycles in the $G_2$ form $\Phi$ (collecting the metric deformations) and in the M-theory 3-form potential $C$:
\begin{equation}
	C + i \, \Phi = i T_I \phi^I, \qquad I = 1,\ldots,7,
\end{equation}
where
\begin{equation}
	\phi^I =\phi^I_{abc}\, dy^a \wedge dy^b \wedge dy^c, 
\end{equation}
for the triplets
\begin{equation}
	5\, 6\, 11, \quad 8\, 7\, 11, \quad 10\, 9\, 11, \quad 6\, 9\, 7, \quad 6\, 8\, 10, \quad 5\, 7\, 10, \quad 5\, 8\, 9,
\end{equation}
We give the list of surviving fluxes in Table~\ref{tab:survflux}, where $a_I, b_I, c_I$ label different indices in the I-th triplet, and by $i_I, j_I, k_I, l_I$ the dual indices in the same triplet.

\begin{table}
	\renewcommand{\arraystretch}{1.4}\addtolength{\tabcolsep}{-1pt}%
\begin{center}
\rowcolors{1}{white}{gray!15}
\begin{tabular}{|c|c|c|}
\hline
	Representation & flux name & d.o.f.s  \\\hline\hline
 $\textbf{1}_{\bf{+7}}$ & $g_{567891011}$ & 1\\
$\overline{\mathbf{35}}_{\bf{+5}}$ & $ g_{i_I j_I k_I l_I}$ & 7\\
 $\mathbf{28}_{\mathbf{+3}}$ & $\tau_{a_I b_I}{}^{c_I}, \tau_{b_I c_I}{}^{a_I}, \tau_{c_I a_I}{}^{b_I}$ & 21 \\
 $\mathbf{7}_{\mathbf{+3}}$ & --  & 0 \\
$\mathbf{\overline{28}}_{\mathbf{+1}}$ & $\xi^{ii}, \quad i=5,\ldots, 11$ & 7 \\
$\mathbf{\overline{21}}_{\mathbf{+1}}$ & -- & 0 \\
$\overline{\mathbf{224}}_{\mathbf{+1}}$ & $R_{i_I}^{j_Ik_Il_I}$, $R_{j_I}^{k_Il_Ii_I}$, $R_{k_I}^{j_Il_Ii_I}$,$R_{l_I}^{i_Ij_Ik_I}$&28 \\
$\mathbf{28_{-1}}$ & $6! \, \theta_{ii} = \epsilon_{i m_1 \ldots m_6}\theta_{i}{}^{m_1 \ldots m_6}, \quad i=5,\ldots,11$ & 7 \\
$\mathbf{21_{-1}}$ & -- & 0 \\
$\mathbf{224_{-1}}$ & $f^{i_I,i_I a_I b_I c_I}$, $f^{j_I,j_I a_I b_I c_I}$, $f^{k_I,k_I a_I b_I c_I}$, $f^{l_I,l_I a_I b_I c_I}$ & 28 \\
$\overline{\mathbf{140}}_{\mathbf{-3}}$ & $Q^{a_I b_I, a_I b_I i_I j_I k_I}$, $Q^{c_I b_I, c_I b_I i_I j_I k_I}$, $Q^{a_I c_I, a_I c_I i_I j_I k_I}$ & 21 \\
$\overline{\mathbf{7}}_{\mathbf{-1}}$ & -- &0 \\
$\mathbf{35_{-5}}$ & $h^{i_I j_I k_I l_I, 567891011}$ & 7 \\
$\mathbf{1_{-7}}$ & $\tilde{g}^{567891011,567891011}$ & 1 \\\hline
\end{tabular}
\end{center}
\caption{Fluxes surviving the G$_2$ invariant truncation given in (\ref{orbifold}).}
\label{tab:survflux}
\end{table}

The scalar potential of the resulting models can be given in a fully geometric form in terms of a superpotential defined as the integral of the fluxes in Table~\ref{tab:survflux} over the internal space, completing the volume 7-form by using the $G_2$-invariant form $\Phi$ and the M-theory 3-form potential $C$.
For instance, geometric fluxes appear in
\begin{equation}
	W = \int_{X_7}\left(g_7 + (C + i \Phi) \wedge g + \frac12 \, (C + i \Phi) \wedge \tau \cdot (C + i \Phi) \right),
\end{equation}
where $\tau$ acts as a differential
\begin{equation}
	\tau \cdot \phi^I = \tau \cdot dy^a \wedge dy^b \wedge dy^c\; \phi^I_{abc} = \frac12 \,\tau_{[de}^{a} \phi^1_{bc]a} \; dy^d \wedge dy^e \wedge dy^b \wedge dy^c .
\end{equation}
We can construct a similar expression for the other fluxes.
For instance
\begin{equation}
	\phi^I \cdot \xi \cdot \phi^J = dy^a  \wedge dy^b \wedge dy^c \wedge dy^d \left(\phi^I_{abi}\xi^{ij} \phi_{jcd}^J\right),
\end{equation}
and the superpotential contribution is
\begin{equation}
	\int_{X_7} (C+ i\,\Phi) \wedge [(C+ i\,\Phi)\cdot \xi \cdot (C+ i\,\Phi)].
\end{equation}

It is straightforward to see that each of the previous fluxes contributes to the superpotential by contracting $(C+ i\,\Phi)^n$, where $n$ is given by 7 minus the charge under the $O(1,1)$ classifying the representations.
This also means that the superpotential has charge 7 and the scalar fields have charge 2.
The number of independent flux components is also equal to the number of the corresponding independent combinations of the 7 moduli.
For instance, $g_4$ and $h^4$ have 7 independent components and indeed they come together with the 7 combinations $T_I$ and $T^6$ where we remove $T_I$, $\tau$ and $Q$ have a total of 21 components, just like $T^2$ and $T^5$. Finally $T^3$ and $T^4$ have 35 independent combinations, which come multiplied by the $28+7$ $R$ and $\xi$ or $f$ and $\theta$ fluxes.

Summarizing, the results is the sum of
\begin{equation}
	\int_{X_7}(C+ i\,\Phi) \wedge ((C+ i\,\Phi)\cdot R \cdot (C+ i\,\Phi)),
\end{equation}
for the R-flux, with
\begin{equation}
	 \phi^I\cdot R \cdot \phi^J = dy^a \wedge dy^b \wedge dy^c \wedge dy^d \; R_a^{mnp} \phi^I_{bcm} \phi^J_{dnp},
\end{equation}
for the $f$-flux
\begin{equation}
	\int_{X_7}(C+ i\,\Phi) \wedge (f \cdot (C+ i\,\Phi)^3),
\end{equation}
where
\begin{equation}
	f \cdot (\phi)^3= dy^a \wedge dy^b \wedge dy^c \wedge dy^d \; f^{j,imnp} \phi^I_{jab} \phi^J_{cd[i}\phi^K_{mnp]},
\end{equation}
for the $\theta$-flux
\begin{equation}
	\int_{X_7}(C+ i\,\Phi) \wedge (\theta \cdot (C+ i\,\Phi)^3),
\end{equation}
where
\begin{equation}
	\theta \cdot (\phi)^3= dy^a \wedge dy^b \wedge dy^c \wedge dy^d \; \theta_a^{i_1\,\ldots i_6} \phi^I_{i_1 i_2 i_3} \phi^J_{i_4 i_5 b}\phi^K_{i_6 cd},
\end{equation}
for the $Q$-flux
\begin{equation}
	\int_{X_7}(C+ i\,\Phi) \wedge (Q \cdot (C+ i\,\Phi)^4),
\end{equation}
where
\begin{equation}
	Q \cdot (\phi)^4= dy^a \wedge dy^b \wedge dy^c \wedge dy^d \; Q^{mn,i_1\,\ldots i_6} \phi^L_{mna} \phi^I_{i_1 i_2 i_3} \phi^J_{i_4 i_5 b}\phi^K_{i_6 cd}
\end{equation}
and for the $h$-flux
\begin{equation}
	\int_{X_7}(C+ i\,\Phi) \wedge (h \cdot (C+ i\,\Phi)^5),
\end{equation}
where
\begin{equation}
	h \cdot (\phi)^5= dy^a \wedge dy^b \wedge dy^c \wedge dy^d \; h^{mnpq,i_1\,\ldots i_7} \phi^I_{i_1 i_2 i_3} \phi^J_{i_4 i_5 a}\phi^K_{i_6 m b} \phi^L_{i_7 np} \phi^M_{qcd}.
\end{equation}


\end{document}